\documentclass{amsart}
\usepackage{cite}
\usepackage{amsmath,amssymb,amsfonts,mathtools}
\usepackage{amsthm}
\usepackage{algorithmic}
\usepackage{graphicx}
\usepackage{textcomp}
\usepackage{xcolor}
\usepackage{framed}
\usepackage{ulem}
\usepackage{soul}
\usepackage{changes}
\usepackage{amsmath}
\usepackage{mdframed}
\usepackage{hyperref}
\usepackage{epstopdf}
\usepackage{todonotes}
\usepackage{comment,xcolor}
\usepackage{bm}
\usepackage{mathtools}
\usepackage[mathscr]{euscript}

\usepackage[bbgreekl]{mathbbol}
\usepackage{amsfonts}
\DeclareSymbolFontAlphabet{\mathbb}{AMSb}
\DeclareSymbolFontAlphabet{\mathbbl}{bbold}

\newcommand{\calC}{{\mathcal{C}}}

\newcommand{\calF}{{\mathcal{F}}}

\newcommand{\calK}{{\mathcal{K}}}
\newcommand{\calL}{{\mathcal{L}}}
\newcommand{\calU}{{\mathcal{U}}}
\newcommand{\calM}{{\mathcal{M}}}
\newcommand{\calR}{{\mathcal{R}}}

\newcommand{\calZ}{{\mathcal{Z}}}

\newcommand{\calW}{{\mathcal{W}}}

\newcommand{\cF}{\mathcal{F}}

\newcommand{\sH}{\mathcal{H}}

\newcommand{\Range}[1]{{\mathscr{R}}(#1)}
\newcommand{\Null}[1]{{\mathscr{N}}(#1)}
\newcommand{\Ext}{{\mathscr{E}}}
\newcommand{\Inj}{{\mathcal{I}}}

\newcommand{\KK}{{\mathbb{K}}}

\newcommand{\RR}{{\mathbb{R}}}

\newcommand{\XX}{{\mathbb{X}}}
\newcommand{\YY}{{\mathbb{Y}}}

\newcommand{\UU}{{\mathbb{U}}}
\newcommand{\ZZ}{{\mathbb{Z}}}
\newcommand{\Knl}{\mathfrak{K}}
\newcommand{\Rnl}{\mathfrak{R}}
\newcommand{\Unl}{\mathfrak{U}}

\newcommand{\knl}{{k}}
\newcommand{\rnl}{{r}}

\newcommand{\sW}{\mathcal{W}}
\newcommand{\sR}{\mathcal{R}}
\newcommand{\sK}{\mathcal{K}}

\newcommand{\vU}{{\bm{\calU}}}
\newcommand{\vZ}{{\bm{\calZ}}}
\newcommand{\vH}{{\bm{\sH}}}
\newcommand{\vR}{{\bm{\calR}}}
\newcommand{\vK}{{\bm{\calK}}}
\newcommand{\vW}{{\bm{\calW}}}

\newcommand{\Pwr}{\mathcal{P}}

\newtheorem{theorem}{Theorem}
\newtheorem{corollary}{Corollary}

\newtheorem{example}{Example}
\newtheorem{remark}{Remark}
\newtheorem{defn}{Definition}

\newcommand{\ipvK}[2]{\left\langle#1,#2\right\rangle_{\vK}}
\newcommand{\ipRd}[2]{\left\langle#1,#2\right\rangle_{\mathbb{R}^n}}

\newcommand{\iptr}[2]{\left\langle#1,#2\right\rangle_{\rm tr}}

\newcommand{\normRd}[1]{\left\|#1\right\|_{\XX}}

\begin{document}
\title[Functional Uncertainty Classes, Nonparametric Adaptive Control]{Functional Uncertainty Classes for Nonparametric Adaptive Control:\\   the Curse of Dimensionality}
\author[Andrew J. Kurdila, et. al.]{Haoran Wang, Shengyuan Niu, Henry Moon, Ian Willebeek-LeMair, \\Andrew J. Kurdila, Andrea L'Afflitto, and Daniel Stilwell}
\thanks{This work is supported in part by the Office of Naval Research (ONR) under the grant number N00014-24-1-2267.}

\maketitle
\begin{abstract}
    This paper derives a new class of vector-valued reproducing kernel Hilbert spaces (vRKHS) defined in terms of operator-valued kernels for the representation of functional  uncertainty arising in nonparametric adaptive control methods. These are referred to as maneuver or trajectory vRKHS $\vK_\calM$ in the paper, and they are introduced to address the curse of dimensionality that can arise for some  types of nonparametric adaptive control strategies.  The  maneuver vRKHSs are derived based on the structure of a compact, $\ell$-dimensional, smooth Riemannian manifold $\calM$ that is regularly embedded in the  state space $\XX\triangleq \RR^n$, where $\calM$ is assumed to approximately  support the ultimate dynamics of the reference system to be tracked. To  achieve an ultimate  target tracking error $\epsilon>0$ that satisfies 
    \[
    \limsup_{t\to \infty}\|x(t)-x_r(t)\|_\XX < \epsilon, 
    \]
    where $x(t)$ is the controlled state and $x_r(t)$ is the reference trajectory to be tracked,  general strategies for constructing the operator-valued kernel  of the maneuver space $\vK_\calM$ are described.   In these methods  the realizable adaptive  controller  is defined in terms $N$  centers $\Xi_N\subset \calM$ of that define a finite dimensional subspace to approximate the uncertainty. In the sharpest bounds derived in the paper, the number of centers $N$ needed to satisfy the above $\epsilon$-target tracking error   scales like 
     \[
     N\triangleq N(\Xi_N,\calM) \sim \frac{1}{\epsilon^{\ell/\bar{s}}}.
     \]
     In this equation,  $\bar{s}>0$ is an integer smoothness index that depends on the regularity of the maneuver space $\vK_\calM$ and $\ell$ is the dimension of the manifold $\calM$. 
\end{abstract}
\section{Introduction}
\subsection{Motivation}
Over the past decade there has been a notable effort to formulate a theory and develop algorithms in adaptive control that are \textit{nonparametric}, as opposed to the  \textit{parametric} strategies that are so well-understood  and  documented  in the now classical references such as \cite{narendraBook,taoBook,sastrybook,ioannouBook,farrellBook,lavretskyBook,krsticBook,slotine1991applied}. The need, as well as the utility, of a general nonparametric adaptive control theory has been  discussed early on in \cite{kurdila2013adaptive,rosenfeld2015state,chowdhary2015bayesian},
also in the recent papers \cite{boffi2022nonparametric,lederer2021uniform,gahlawat2020rl1}, and  by  the authors in \cite{kurdilaBook}.

In view of the growth in  importance of methods of statistical and machine learning theory over the past decade, it is perhaps unsurprising that essentially all of the above work in formulating nonparametric adaptive control has analyzed the problem when the functional uncertainty is contained in reproducing kernel Hilbert spaces (RKHSs) $\sK$ that contain real-valued functions or vector-valued RKHSs (vRKHSs) $\vK$ 
that contain real, vector-valued functions. References \cite{liu2018gaussian,boffi2022nonparametric,chen2022gaussian,umlauft2017feedback,umlauft2018uncertainty,umlauft2019feedback,jagtap2020control,beckers2017stable,beckers2019stable,fan2020bayesian,dhiman2021control,gahlawat2020l1,helwa2019provably,konig2021safe,capone2019backstepping,westenbroek2020feedback,sinha2022adaptive,jiao2022backstepping,lederer2020training,arabi2019neuroadaptive,sinha2022adaptive,sforni2021learning,nayyer2022passivity,chen2019gaussian,joshi2018adaptive,grande2014experimental,ignatyev2023sparse,chowdhary2015bayesian,chang2017learning,grande2016online,wajid2022formal,care2023kernel,he2022adaptive}  all discuss adaptive control strategies where the functional uncertainty error is analyzed in a stochastic setting based on techniques of Gaussian processes (GPs). On the other hand, the  recent text \cite{kurdilaBook} consolidates results in \textcolor{red}{\cite{kurdila_and_lafflitto_papers} } and gives a detailed \textit{deterministic}  analysis by the authors of nonparametric adaptive control for functional uncertainties. 

These latter methods can be distinguished from the GP control approaches above in many ways, but one of  the most fundamental is that they rely on describing an  ideal  deterministic control system  that is  a   limiting  distributed parameter system (DPS) that evolves in a generally infinite dimensional state space. Realizable controllers are always  constructed from consistent finite dimensional approximations of the deterministic  limiting DPS. Another significant difference is that the GP-based adaptive control methods rely on stochastic approximations of the uncertainty $f$ derived by random sampling of the  Gaussian process  $f$, thereby obtaining ultimate tracking  error bounds that hold with high probability. On the other hand, the deterministic approaches employ approximations  for which deterministic ultimate tracking error bounds are determined using  various definitions of the power function, the many zeros theorems, or deterministic greedy approximation methods. \cite{kurdilaBook}

This paper addresses one of the more recent  challenges encountered in formulating deterministic nonparametric adaptive control theory for such functional uncertainty classes: we analyze the  \textit{computational complexity} of controller performance guarantees for ultimate tracking error. Recall that in the fields of information based complexity theory \cite{traub1998complexity,traub2003information}, statistical  learning theory \cite{gyorfi2002distribution}, or approximation theory \cite{devore1998nonlinear}, the computational complexity of an algorithm to approximate an unknown function is a description of the amount of computational work that must be performed to obtain a prescribed accuracy. The corresponding problem for us is the characterization of the amount of computational work required in a nonparametric  feedback control strategy to achieve a prescribed tolerance on the ultimate  tracking error.  Of the above references, only the very recent work in \cite{boffi2022nonparametric} studies this problem specifically, but  in the stochastic  setting of nonparametric adaptive control via Gaussian processes. In this reference GP methods are used to determine the number of centers needed to obtain target tracking error bounds that hold with high probability. It is also noteworthy that the GP methods in \cite{umlauft2017feedback,umlauft2018uncertainty,umlauft2019feedback,lederer2021uniform,gahlawat2020rl1} or the deterministic methods in \cite{farrellBook,lavretskyBook,kurdilaBook} describe results that could support the further development  of complexity estimates in a stochastic or deterministic setting, but have not as of  yet treated the characterization of computatational complexity of controllers \textit{per se}.    

We review in more detail some of the properties of stochastic GP-based analysis  in Section \ref{sec:GPsreview},  while the foundations of  deterministic approaches are reviewed in Section \ref{sec:powerfunctionmethods}.  The  literature review further explains the novelty and emphasizes the differences among the deterministic approaches  introduced in this paper and these stochastic GP-based  methods. 

\subsection{The Problem Statement}
To motivate the research described in this paper, we initially consider the model problem
\begin{align}
\dot{x}(t)=Ax(t) + B(u(t) + f(x(t)) ) + \gamma(t) \label{eq:model1}
\end{align}
where the state $x(t)\in \XX\triangleq \RR^{n}$, the control $u(t)\in \UU\triangleq \RR^{m\times 1}$, the system matrix $A\in \RR^{n\times n}$, the control influence matrix $B\in \RR^{n\times n}$, and the unmatched uncertainty $\gamma(t)\in \XX$. Throughout this paper we assume that the matched uncertainty resides in some selected RKHS of vector-valued functions, referred to simply as a vRKHS, so that  $f\in \vK\triangleq \vK(\XX,\UU)$, where  $\vK$ is defined in terms of an operator-valued  kernel $\Knl(x_1,x_2)\in \calL(\UU)$ for all $x_1,x_2\in \XX$. A discussion of the most basic, but relevant, properties of operator kernels and vRKHS is given in Section \ref{sec:operatorvRKHS}. During this motivating discussion in this introduction,  we just assume $A$ is Hurwitz, $A$ is known, $B$ is known, and $\gamma(t)\equiv 0$. These assumptions are eliminated later in the detailed  analysis of the closed loop control schemes in Section \ref{sec:nonparametric_and_jackson_inequalities} and in the consideration of concrete examples in Section {\ref{sec:numerical_examples}.}

 The goal is to determine an adaptive  feedback controller $u(t)\triangleq \mu(t,x(t))$   to drive the system state to track some reference trajectory $t\mapsto x_r(t)\in \XX$, so that 
\[
\lim_{t\to \infty} \|x(t)- x_r(t)\|_\XX = 0.
\]
If such an ideal performance cannot be achieved, as is usual in robust modifications of adaptive control, we seek to establish uniform ultimate boundedness 
\[
\limsup_{t\to \infty} \|x(t)-x_r(t)\|_\XX \leq \epsilon
\]
for some suitably small controller performance bound $\epsilon>0$.

Suppose we are studying the performance of controllers for some set of initial conditions $x_0\in \Omega_0 \subseteq \XX$, and we  choose centers $\Xi_N\in \XX$ (as discussed in detail in Section \ref{sec:interpolation}) that define the finite dimensional spaces $\vK_N\subseteq \vK$ that are  used to approximate the nonparametric  uncertainty $f$ and construct realizable controllers. A variety of specific feedback controllers in robust modifications of parametric adaptive control theory, \cite{lavretskyBook,farrellBook} ensure that  the closed loop trajectories remain in the bounded set $\Omega \subset \XX$.   The culimination of the work in the deterministic analyses of methods in  references \cite{kurdilaBook,annualreviews2023_a,annualreviews2023_b} is a collection of performance bounds on adaptive control strategies that have the form 
\begin{align}
    \underbrace{\limsup_{t\to \infty} \|x(t)-x_r(t)\|_{\XX}}_{\text{measure of  controller performance}} \lesssim \underbrace{ R(N)\triangleq R(N;\Omega_0,\Omega)}_{\text{measure of offline approximation error} } \label{eq:generalformerror}
\end{align}
\textit{for all uncertain systems} in Equation \ref{eq:model1} such that the unknown function $f\in \calC \subset \vK$ for  $\calC$ a suitable functional uncertainty class contained in the vRKHS $\vK$. In this equation, $R(N)$ is a rate function that represents the rate of convergence of \textit{(offline) approximations} that hold for any  $f \in \calC$ using $N$ centers, and it satisfies $R(N)\to 0$ as $N\to \infty$. These performance bounds hold for all the coordinate instantiations for consistent approximations  of a particular nonparametric control scheme ``bundled together.''  {\cite{kurdilaBook}}. Here we see that the general structure of the \textit{online inequality bounds on the controller performance} are expressed in terms of \textit{offline bounds on  the quality  of approximations} over all the functions in the uncertainty class $\calC \subset \vK$ in the vRKHS $\vK$. 

Among the many adaptive control schemes and associated adaptive controller performance bounds summarized in \cite{kurdilaBook}, or for the references above on Gaussian processes, the results for deadzone methods are among the simpler  to state. We summarize them here as exemplars of the qualitative nature of such nonparametric approaches.  It has been shown \cite{kurdilaBook} that, under suitable hypotheses and choices of the vRKHS $\vK$, a properly designed deadzone method can yield  controller performance bounds that have the form 
\[
\limsup_{t\to \infty}\|x(t)-x_r(t)\|_\XX \lesssim \sup_{x\in \Omega}\|E_x(I-\bm{\Pi}_N)f\|_\UU \lesssim O(h_{\Xi_N,\Omega}^s)
\]
for all systems in which the functional uncertainty $f$ resides in the uncertainty class $\mathcal{C} \subset \vK$. Here $E_x:f\mapsto f(x)$ is the evaluation operator at $x\in \XX$,  $\bm{\Pi}_N$ is the $\vK$-orthogonal projection onto  the finite dimensional space $\vK_N$,  $s>0$ is a \textit{smoothness parameter} that measures how regular the functions are in the vRKHS $\vK$, and $h_{\xi_N,\Omega}$ is the fill distance of the samples $\Xi_N\subset \Omega$, which is defined as 
\[
h_{\Xi_N,\Omega}\triangleq \sup_{x\in \Omega} \min_{\xi_i\in \Xi_N} \|x-\xi_i\|_\XX. 
\] 
We refer in \cite{kurdilaBook} to the class of adaptive control methods that have a bound such as this as \textit{asymptotically approximation theory optimal}, or AAO for short. This terminology reflects the fact that the ultimate controller tracking performance is bounded above by the maximum pointwise  error of the best approximation $\bm{\Pi}_N f$ from $\vK_N$ of the uncertainty $f\in \vK$. 
Establishing when adaptive control schemes are AAO is a good starting point for comparisons among alternative algorithms. In this sense AAO methods are understood as types of ideal nonparametric adaptive control methods. 

The above bound is quite general and holds for a vast collection of choices of the reproducing kernel that defines the vRKHS $\vK$. It constitutes an important step  toward establishing a useful general theory of nonparametric adaptive control since it  enables the systematic study of a  problem that has  as of yet largely gone unanswered in the conventional form of  parametric adaptive control theory. Specifically, the nonparametric  framework above enables a careful assessment of the following question:

\begin{center}
\parbox[t]{3.0in}{\textit{How does the regularity or smoothness of the uncertainty affect the ultimate tracking performance of a nonparametric  adaptive control method?}}
\end{center}
\medskip 

\noindent The fact  that the bound above is explicit in the regularity $s$ of the uncertainty class enables the study of the question above in precise terms, which we are only just beginning to understand as a  research community.   

At the same time, the nonparametric theory and framework enables the consideration of still another question in this paper, one that is more commonly encountered in studies pertaining to statistical and machine learning theory, or in approximation theory. 
In view of  the nonparametric adaptive controller  bound above, in this paper  we seek to answer the related question below:

\medskip 
\begin{center}
\parbox[t]{3.0in}{\textit{What is the computational complexity of obtaining an ultimate tracking error target $\epsilon>0$ for a nonparametric adaptive control method?}}
\end{center}
\medskip 

\noindent In classical studies of statistical and machine learning theory \cite{scholkopf,williams2006gaussian}, information-based complexity \cite{traub1998complexity,traub2003information}, or classical approximation theory \cite{devore1998nonlinear,dl1993}, this is a standard question regarding offline approximation of functions.  

Despite the fact that the definition of an AAO method described above gives a nice characterization of the ultimate tracking error, and begins to explain how it depends on the regularity of the uncertainty, it can suffer from the ``curse of dimensionality,'' depending on how it is applied. Suppose  we only have a very vague description of the set $\Omega$ in which the closed loop trajectory lies.  We might know for instance that  it is contained in a set $\Omega\triangleq [a_1,b_1]\times \cdots [a_n,b_n]\subset \XX$ that is a parallelopiped in $\XX\triangleq \RR^n$. In this case, if the centers $\Xi_N\subset \Omega$ that define the space of approximants $\vK_N\subset \vK$ are ``well spread out'' in the set $\Omega$, we  show in Corollary \ref{cor:curse} in Section \ref{sec:nonparametric_and_jackson_inequalities} that the number of centers $N$ dictated  by setting the target accuracy to $\epsilon>0$ in the above controller performance bound scales as follows: 
\begin{equation}
\text{proxy for the computational work}\triangleq {N} \sim O\left ( \frac{1}{\epsilon^{n/s}}\right).  \label{eq:form_complexity}
\end{equation}
This is a classical example   of computational complexity that exhibits the curse of dimensionality in terms of the number of state space variables $n$. The right hand side grows exponentially as the dimension $n$ of the state space $\XX$ of the control problem increases. 

 \textit{This paper uses information specific to the adaptive tracking control problem to construct infinite dimensional vRKHS $\vK_\calM\subseteq \vK$ for the definition of  nonparametric uncertainty classes that improve, or even eliminate, the poor  scaling  noted above as the dimension of the state space increases.}

\subsection{Contrasting Problems of  Control to  Machine Learning Theory}
The philosophical similarity of the above question for nonparametric adaptive control  to the corresponding problem of offline  function approximation in  machine learning and nonparametric regression  is obvious, but there are some 
 features   that are unique to the control setting. It is worth emphasizing these differences.  Firstly, it is more or less standard to approximate solutions of  regression problems for an unknown  $f$  assuming the availability   of noisy or error-ridden samples $\{(x_i,y_i)\}_{i=1}^M$ of the input-output behavior of $f$, with $\{(x_i,y_i)\}_{i=1}^M\approx \{(x_i,f(x_i))\}_{i=1}^M$. However, in the setting of control problems it is most frequently the case  that $f$ is not accessible to direct measurements. The  best for which  we can usually hope is that the system is instrumented to make online measurements of all of the states $x(t)$, and even this may not be possible.  As an example in vehicle control, for either marine or flight vehicles, the uncertainty $f$ is ordinarily understood to represent complicated, nonlinear aerodynamic or hydrodynamic loads. Actual controlled vehicles are (almost) never instrumented to yield measurements of such complex input-output behavior  in real-time.  This is true even if the vehicles  are experimental or otherwise are examples of  specialized test vehicles.

 Another important difference is that in feedback control problems like the one above,  it is not ordinarily the  goal to estimate the uncertainty $f$ as an end unto itself.  If we could measure the input/output behavior of $f$ in real-time, and if we subsequently use such measurements to  estimate $f$ accurately, then it is rather straighforward to construct good controllers that give excellent measures of performance as described above.  But, as carefully laid out in all the classical texts above on parametric adaptive control,  good approximations constitutes only  a ``sufficient condition''  to enable or design good controllers. And,  in view of the comments above about the inaccessibility of $f$, there are many popular control strategies that yield excellent, rigorous guarantees of controller tracking error performance without corresponding   rigorous proofs of the convergence of the error of  estimates  of  the uncertainty.

 As  we expand upon in the next section, one of the critical qualitative features of closed loop tracking control systems is that the deterministic  states  often  naturally cluster in  smaller and smaller neighborhoods of some limiting compact set, which can be highly irregular.   Of course, the  controlled system is typically designed to converge to the neighborhood of some small subset of Lebesgue measure zero in the state space, like a fixed point or like a one dimensional compact submanifold. 
 
 Finally, data-driven adaptive control schemes often build controllers based on samples collected along the state trajectory. But the  samples along a deterministic  state  trajectory are typically neither independent nor identically distributed. This is a stark contrast from the  starting hypotheses of many of the standard learning and regression problems analyzed using Gaussian processes that assume samples are generated by a random walk determined by an unknown probability measure  over some fixed, known subset.  

\bigskip 
\begin{figure}[h!]
\label{fig:AUV}
\begin{tabular}{ccc}
\includegraphics[width=0.52\textwidth]{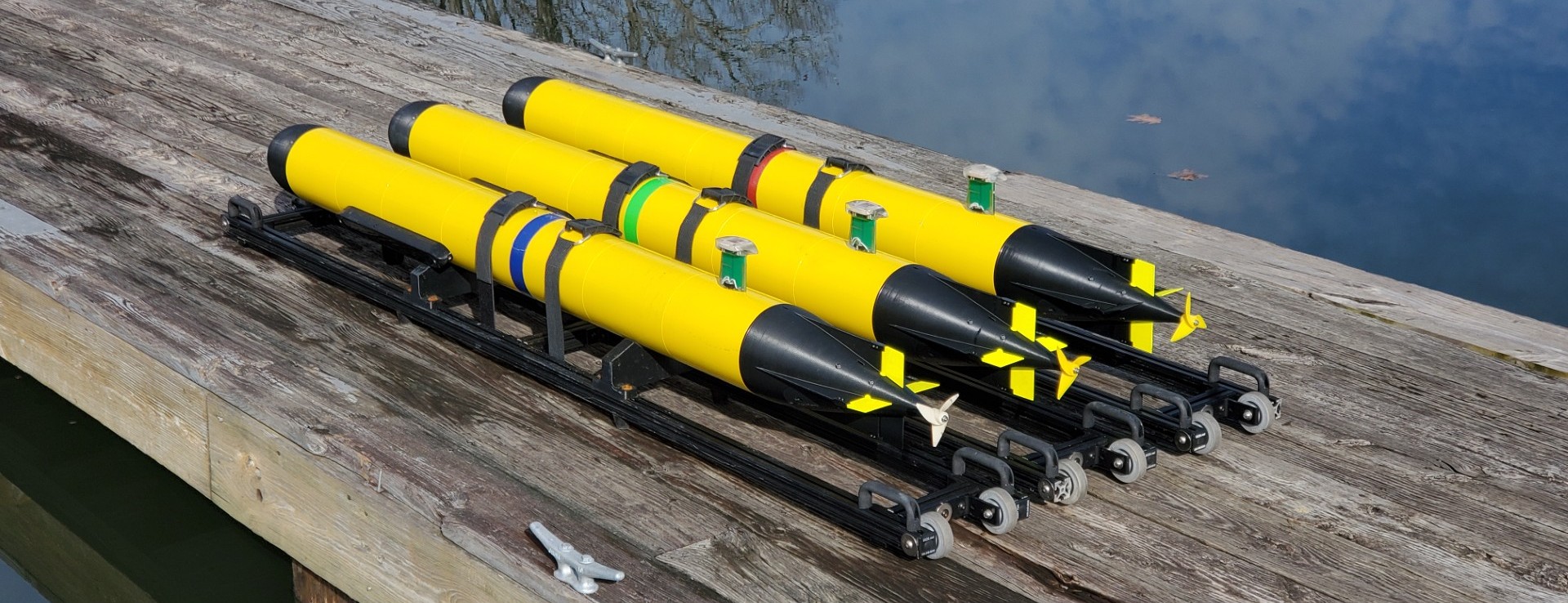} 
&
\includegraphics[width=0.4\textwidth]{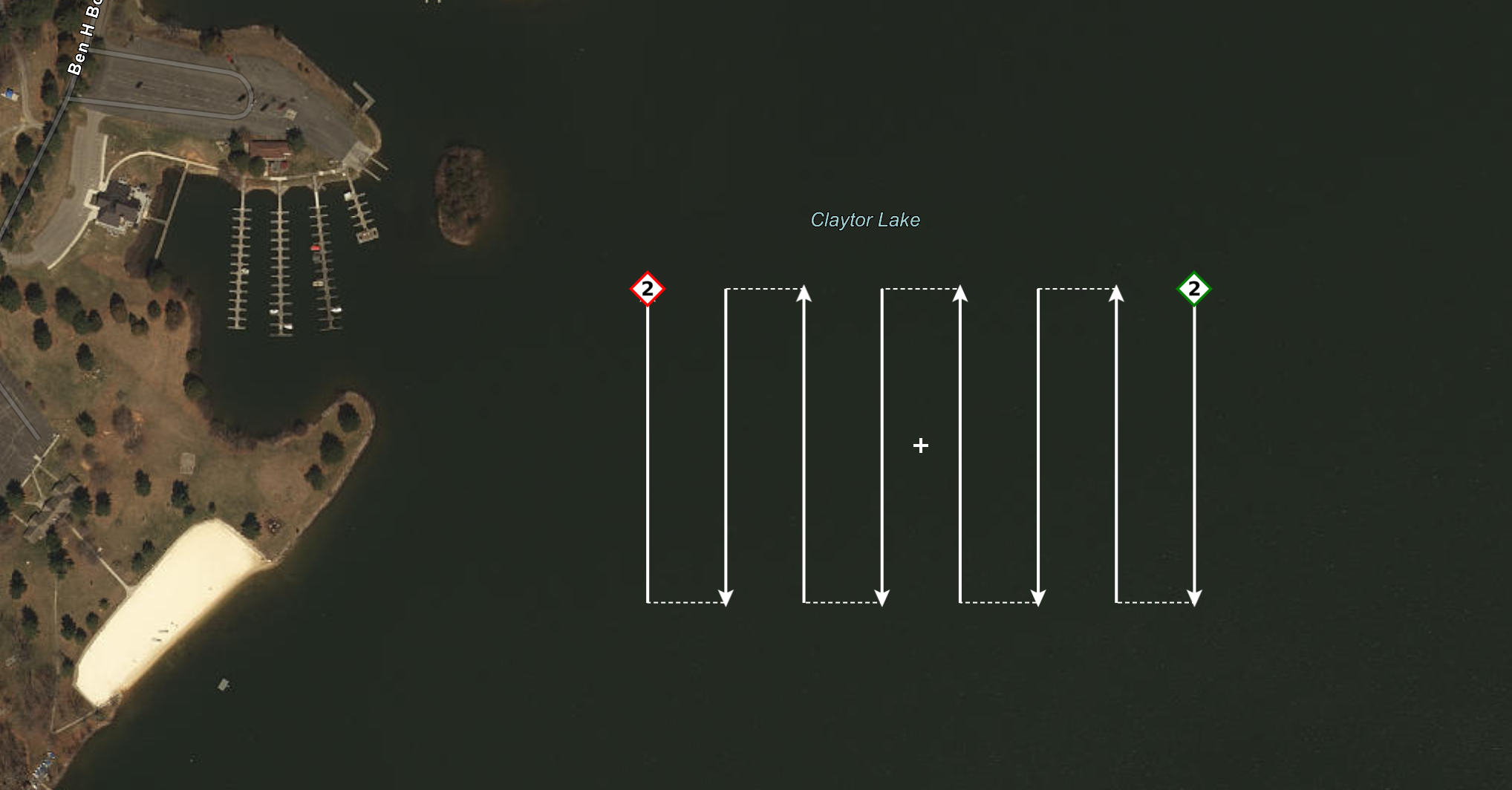} 
\\
(a) Virginia Tech 690 AUV
& 
(b) User commanded survey profile
\end{tabular}
\includegraphics[width=1.0\textwidth]{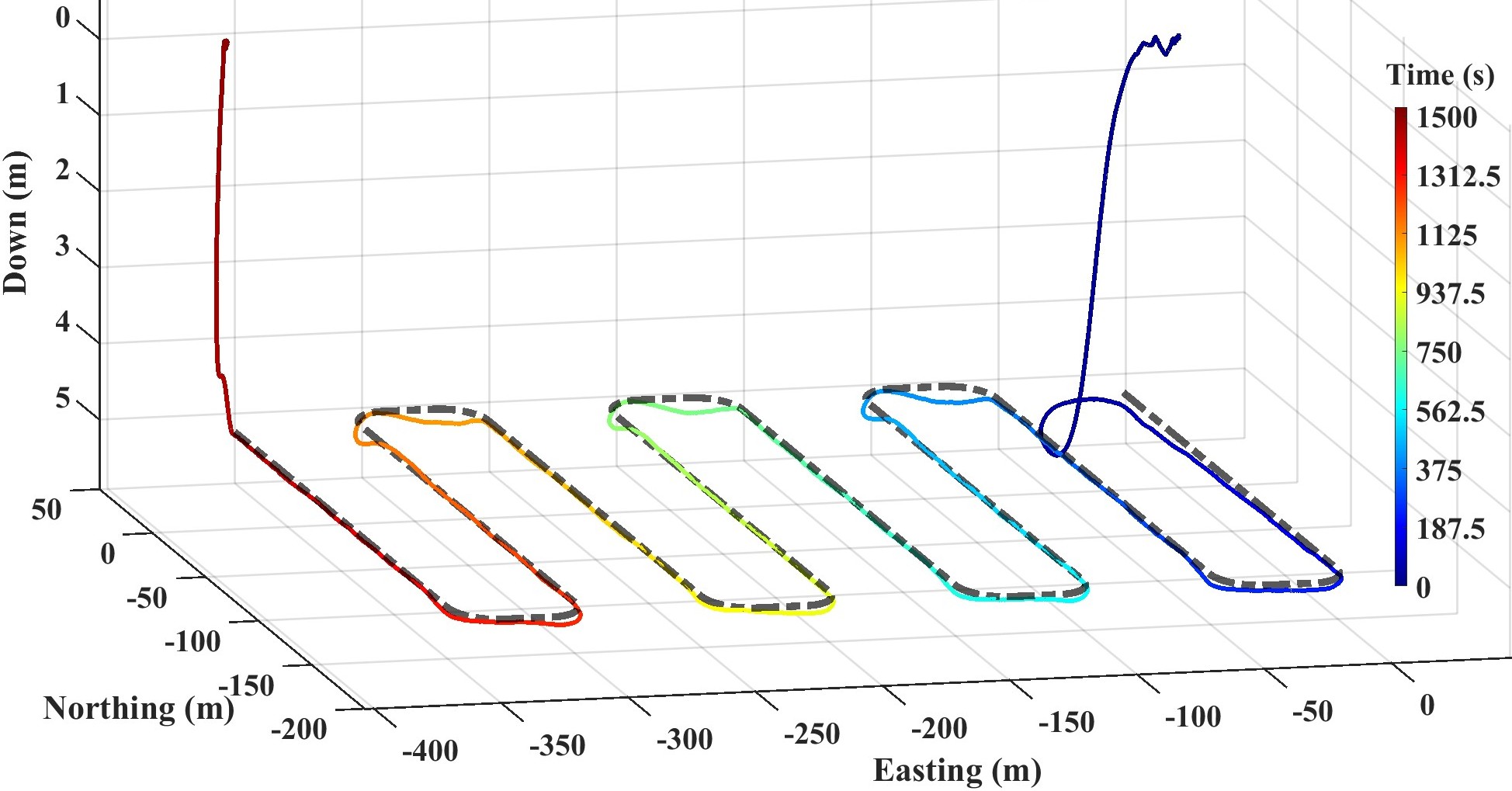}
\\
(c) North-East-Down (NED) coordinates of the AUV during the survey action with origin at the mission start position
\caption{ }
\end{figure}

 \begin{example}[Example for Autonomous Underwater Vehicles (AUVs)]
     As a concrete example to provide further motivation for the work in this paper, we consider the control of autonomous underwater vehicles shown in Figure (a). AUVs are commonly tasked with surveying extensive areas to collect environmental data, such as bathymetry, salinity, and temperature, or to search for specific targets of interest. To do so, we command a survey profile in a lawnmower-pattern as illustrated in Figure (b). In this particular example, the AUV executes an initial dive action to a depth of 5 meters followed by a survey trajectory consisting of eight 200 meters long swaths spaced 50 meters apart. Figure (c) depicts the output of the AUV's inertial navigation algorithm, indicating the estimated position of the AUV color-coded by time, and the desired trajectory as a black dashed line. The full order state models for such systems are complex systems of nonlinear ODEs often containing difficult-to-model hydrodynamic coefficients. See \cite{njaka2022} for the detailed study of the Reynolds-Averaged Navier-Stokes CFD modeling for such AUVs, as well as an analysis of associated approximations for the ODEs governing the motion of the AUVs. The full set of approximating ODEs evolve in a state space with $n=12$. If we only utilize the knowledge that the closed loop controlled trajectories evolve in some hypercube, the complexity estimates above make detailed approximations for centers well-distributed  over the hypercube intractable. However, as is often the case in practical control algorithms, initial controllers are often available that achieve some level of nominal performance that we wish to improve. As depicted in Figure (c), such controllers typically generate trajectories that eventually  reside in the neighborhood of some desired trajectory. The aim of this paper is to use such information to tailor the definition of uncertainty classes for  nonparametric adaptive controllers  to achieve performance guarantees that do not suffer the curse of dimensionality. Intuitively, we desire the ``largest uncertainty classes' consistent with the computational limits associated with approximating the uncertainty. 
 \end{example}

\subsection{The Theoretical Open Questions and Contributions}
Several factors in the formulation of the design of the controller affect the form of the inequality in Equation \ref{eq:generalformerror}. These include the choice of the hypothesis space $\vK$, uncertainty class $\calC\subset \vK$, the set $\Omega$ that contains the closed loop trajectories, the set of centers $\Xi_N\subset \XX$ that define $\vK_N$,  and the set $\Omega_0$ of initial conditions. All of the contributions of this paper derive ways to modify the choice of uncertainty classes $\calC\subset \vK$ to address situations when the complexity estimates scale poorly as the dimension $n$ of the state space increases.

The overall approach taken in this paper differs fundamentally from the usual strategies encountered in statistical and machine learning theory, information-based complexity theory, or approximation theory.  \textit{We show how uncertainty classes can be tailored to the type of information that is typically  available in tracking control problems. We use the knowledge of the ultimate structure  of the dynamics of  the reference system to be tracked to define the functional uncertainty classes.}

Specifically, we carry out the following steps in this paper to develop our complexity estimates for ultimate tracking performance.  
\begin{enumerate}
    \item We first  study some  simple cases where the lack of detailed information about the set $\Omega$ containing a closed loop trajectory results in  complexity  estimates for nonparametric methods  derived in \cite{annualreviews2023_a,annualreviews2023_b,kurdilaBook} that exhibit a classic ``curse of dimensionality'' in the form of Equation \ref{eq:form_complexity}. This is described precisely in Corollary \ref{cor:curse}.   We also explain how this  case can be interpreted as applying also to some other  classical parametric control approaches as in \cite{lavretskyBook,farrellBook}. 
    \item  We derive smaller, but generally infinite dimensional, functional uncertainty subspaces  $\vK_{\calM}\subseteq \vK$ that enable computational complexity estimates that exhibit better scaling. The construction or definition of the hypothesis space $\vK_\calM$ depends on a smooth $\ell$-dimensional manifold $\calM\subset \XX$ that approximately supports the long term dynamics of the reference trajectories $t\to x_r(t)\in \XX$ to be tracked. The manifold $\calM$ could also be defined to represent the ultimate dynamics of  families of reference systems in a general control design problem.  Since we are aiming at techniques that can be applied to vehicle tracking control we refer to the hypothesis spaces $\vK_\calM\subset \vK$ as \textit{maneuver or trajectory vRKHS}.  The intuition behind this approach is that we choose to spend our ``approximation budget'' consistent with our knowledge of where the state of the  reference system ultimately accumulates,  so as to improve the ultimate tracking error.  
    \item We show that  for these new  nonparametric uncertainty classes the corresponding robust performance bounds for uncertainty in the maneuver space $\vK_\calM$ do not suffer from a curse of dimensionality.   Instead, we argue that to achieve an ultimate tracking error $\epsilon>0$ in 
    \[
    \limsup_{t\to \infty}\|x(t)-x_r(t)\|_\XX \sim \epsilon, 
    \]
    the performance bound for all systems with the uncertainty $f\in \vK_\calM$  yield computational  complexity estimates that scale like 
    \begin{align}
    N \sim \frac{1}{\epsilon^{\ell/\bar{s}}}\label{eq:complexity_M}
    \end{align}
    where $\ell$ is the dimension of the embedded manifold $\calM\subset \RR^n$ and $\bar{s}$ is the \textit{reduced smoothness}. This result is described rigorously in Corollary \ref{cor:diagonal_R_M}. It is important to emphasize that the dimension $\ell$ of the submanifold can be understood as a \textit{design variable} and will generally be taken so that $\ell <<n$ where $n$ is the dimension of the state space $\XX\triangleq \RR^n$. The new bound above in Equation \ref{eq:complexity_M} for $f\in \vK_\calM$ and $\Xi_N\subset \calM$ should be compared carefully with that above in Equation \ref{eq:form_complexity}for $f\in \vK$ and $\Xi_N\subset \Omega$. 
\end{enumerate}

We conclude this paper by studying the qualitative behavior of the proposed nonparametric control approaches in numerical studies in Section \ref{sec:numerical_examples}.

\subsection{Relevant Research and Literature}
All of the nonparametric adaptive control approaches mentioned above, for either the methods based on properties of  GPs, or for the  deterministic approaches based on  approximations of a deterministic limiting  DPS, seek to reduce the information that some expert or oracle must supply to choose finite dimensional subspaces for  construction of realizable controllers. In parametric adaptive control methods, the choice of the finite dimensional subspace is made by an oracle and is not considered part of the adaptive control method \textit{per se}.  In nonparametric adaptive control, the choice of the subspaces is considered an important unknown that the adaptive controller must choose or identify. 

\subsection{Methods Based on GPs, IID Samples, and Stochastic Analysis}
\label{sec:GPsreview}
\begin{figure}[h!]
\label{fig:GPsamples}
\begin{tabular}{ccc}
\includegraphics[width=.30\textwidth]{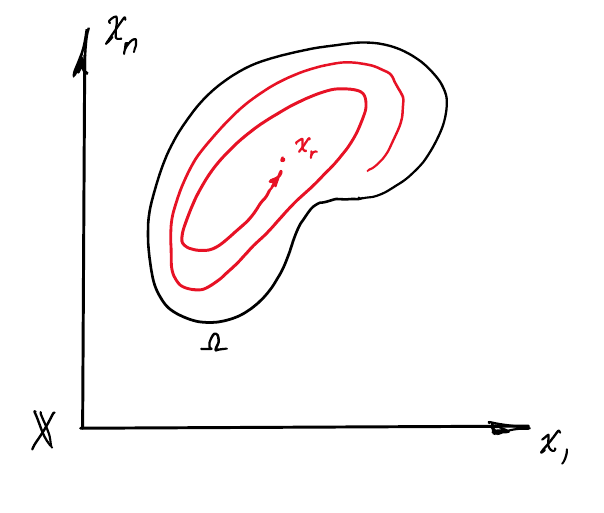}
&
\includegraphics[width=.30\textwidth]{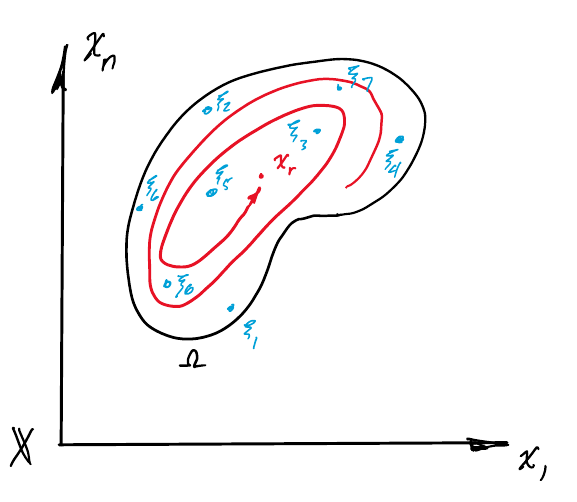}
&
\includegraphics[width=.30\textwidth]{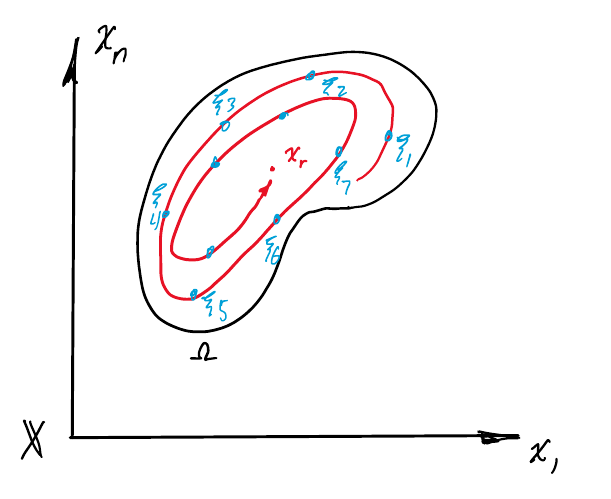}\\
(a) Controlled trajectory 
&
(b) Centers for GP
&
(c) Trajectory-driven centers
\end{tabular}
\caption{This figure shows a typical controlled trajectory and centers selected. Here the controller is designed to drive the system to a fixed target reference state $x_r$. More generally, it is desired to choose the controller to drive the trajectory to track some reference trajectory $t\to x_r(t)$. As shown in (a), as time progresses, the state enters and remains in progressively smaller neighborhoods of the target. Figure (b) above illustrates that there  is no correlation between centers ordinarily chosen in application of GP methods and the state trajectories, at least in the theoretical proofs of convergence.  Figure (c) depicts a family of trajectory-driven centers: they are selected from the positive  orbit $\gamma^+(x_0)=\bigcup_{t\geq 0}x(t).$ }
\end{figure}

For the stochastic approaches its is usually assumed ithat there is measurement process in discrete time that generates samples $\{(x_i,y_i)\}_{i=1}^M$ that approximate the true input/output response $\{(x_i,f(x_i)\}_{i=1}^M$ of the functional uncertainty $f$. Consistent with the theory of GPs, it is always assumed in the adaptive control references above   that the samples  are independent and identically distributed (IID) and are generated by some user-defined probability measure on $\ZZ=\XX \times \YY$. This situation is depicted schematically in Figure \ref{fig:GPsamples}.  

The choice to employ GP methods for adaptive control of deterministic nonlinear ODEs in continuous time is due to their spectacular success in solving a number of important classical offline methods of  statistical and machine learning theory  for  regression problems. These classical problems  use  IID measurements in discrete time that are generated by a random walk over the domain of interest. The philosophy above using GP methods for nonlinear ODEs in continuous time also has a number of distinct advantages.
(1) It allows for the most direct incorporation of well-known features of GP approximation methods based on discrete IID samples in a nonparametric adaptive control strategy. (2) Some detailed bounds on steady-state controller performance have been derived using concentration of measure theorems in $L^\infty(\Omega,\XX)$ for (Lipschitz continuous) kernels.  

Despite these promising attributes there are a few important challenges that must be addressed to continue to extend the overall GP strategy to wider classes of systems and to derive more widely applicable performance bounds. (1) The assumption of an underlying  IID measurement process that samples the input/output response of the matched functional  uncertainty is at best counter-intuitive for deterministic nonlinear ODE systems that evolve in continuous time. One is immediately confronted with describing the physical mechanism by which such IID samples of $f$ are obtained in practice. Also,   some GP approaches assume that the pair $(x_i,\dot{x}_i)\equiv (x(t_i),\dot{x}(t_i))$ are samples of a GP, which is equivalent in many situations to the assumption that samples of the input/output response of the  GP $f$ are available.  Such methods that rely on measurements of the state derivatives $\dot{x}(t)$ must address one of the most universal truisms of experimental methods: measurements of derivatives are inherently noisy, sometimes prohibitively so. Our own experience with simulations of high-dimensional autonomous underwater vehicles in \cite{oesterheldcdc2023} have shown that, perhaps unsurprisingly,   samples (or estimates) of the derivatives can be so  noisy in practice that it makes GP methods based on them problematic.  (2) Another potential challenge is that the the error bounds on steady state performance in the adaptive control methods based on GPs rely explicitly on the availability of concentration of measure inequalities in the space $L^\infty(\Omega,\XX)$. As emphasized by \cite{lederer2019uniform}, these inequalities are rather scarce, although this topic overall has seen more interest very recently in the theory of operator kernels and vRKHS.  It is much more common in statistical and machine learning theory that concentration inequalities are derived or stated in $L^2(\Omega,\XX)$. See for example the references \cite{temlyakov2011greedy,smale2007learning,devore1998nonlinear}. But the plethora of concentration inequalities in $L^2(\Omega,\XX)$ are not directly applicable in applications to adaptive control, as least not how they are used in \cite{lederer2021uniform,kurdilaBook}. See the recent text \cite{kurdilaBook} for a detailed account.  (3) Finally, for any reasonable adaptive controller, we expect that the controller trajectory $t\mapsto x(t)$ gets closer and closer over time to smaller neigborhoods that contains the reference trajectory $t\mapsto x_r(t)$. If we choose a fixed  probability measure over $\Omega$ the defines the random walk used for samples of a GP, it is intuitive that it could be attractive   to redefine the measure used to define the random walk to smaller and smaller neighborhoods of the desired trajectory.  But the description of exactly how such a strategy can be carried out in practice, how the size of the progressively smaller sets are selected,  how the switching of the  probability measures  is triggered,  a careful analysis of the resulting hybrid switched system (having time-dependent dimension over progressively smaller sets), and how this structure enables  rigorous theoretical guarantees on controller performance have not yet  been derived in detail.

\subsection{Methods Based on Power Functions and Deterministic Analysis}
\label{sec:powerfunctionmethods}

In view of the above comments, it seems natural to study deterministic adaptive control strategies for ODEs where the samples are collected along a trajectory.  We call such methods trajectory-driven methods in this paper. 
To get an idea of the origins of the specific contributions described below, we summarize the primary thrust of the  trajectory-driven approaches in 
\cite{annualreviews2023_a,annualreviews2023_b,kurdilaBook}. In these references it is assumed that the matched  uncertainty resides in an functional uncertainty class $\calC\subset \vH$  contained in the general v-RKHS space $\vK=\vK(\XX,\UU)$ that is defined in terms of a (possibly nondiagonal) operator kernel $\Knl(x_1,x_2)\in \calL(\UU)$.
Practical controllers are defined in terms of a finite dimensional subspace $\vK_N\subset \vK$, by using  the $\vH$-orthogonal projection $\bm{\Pi}_N:\vK\to \vK_N$. 
In these references, two types of functional uncertainty classes are introduced, one larger and one smaller, 
\begin{align}
    {C_R}&\triangleq \{f\in \vK \ | \ \|f\|_{\vK} \leq R\} \subset \vK, \label{eq:bigclass}\\
    \calC_{R,\epsilon,N}&\triangleq \left \{ f\in \vK \ | \ \|f\|_{\vK} < R, \|(I-\Pi_N)f\|_\vK \leq \epsilon \right \} \subset {C_R} \subset \vK. \label{eq:smallclass}
\end{align}
We seek controller performance guarantees in the form of Equation \ref{eq:model1} for all  $f\in \calC\subset \vK$, with $\calC$ selected as either $\calC_R$ or $\calC_{R,\epsilon,N}$. 
The symbol $\lesssim$ in the above inequality allows that there can be a constant $C$ on the right hand side, which can depend on the control method, which is not shown for brevity. In this equation the variable $R(N)$ is a \textit{rate function} that satisfies $R(N)\to 0$ as $N\to \infty$ for all functions $f\in \calC\subset \vK$. 

The approaches summarized in \cite{annualreviews2023_a,annualreviews2023_b,kurdilaBook} describe a large collection of individual algorithms that generalize well-known methods in classical parametric adaptive control to the nonparametric setting where functional uncertainty lies in the uncertainty classe $\calC$, which is contained in an  vRKHS $\vK$ or RKHS $\sK$. The guarantees often have the form 
\begin{align*}
    \limsup_{t\to \infty} \|x(t)-x_r(t)\|_\XX &\lesssim \sup_{\xi\in \Omega} \overline{\Pwr}_N(\xi) \|(I-\Pi_N)f\|_\vH, \\
    &\lesssim \sup_{\xi\in \Omega} \overline{\Pwr}_N(\xi) \|f\|_\vH, 
\end{align*}
where $\overline{\Pwr}_N$ is the  power function for the finite dimensional space $\vK_N$ in the v-RKHS $\vK$. It is defined \cite{kurdilaBook} as 
\begin{align*}
\overline{\Pwr}_N(x) & = \sqrt{\|\Knl(x,x)-\Knl_N(x,x)\|}
\end{align*}
where $\Knl(x_1,x_2)$ and $\Knl_N(x_1,x_2)$ for $x_1,x_2\in \XX$ are the operator kernels that define $\vK$ and $\vK_N$, respectively. Once the $N$ centers  $\Xi_N$ are known or selected, this power function can be evaluated since the original kernel  $\Knl$, and therefore $\Knl_N$ are known.  When $f$ is contained in the RKHS $\sK$, or when the vRKHS $\vK$ is a Cartesian product $\vK\triangleq \sK_1 \times \cdots \times \sK_m$, even stronger results can be derived as described in \cite{kurdilaBook}. For some standard choices of kernels, such as the class of Sobolev-Matern kernels or Wendland kernels as described in \cite{wendland} and \cite{schaback94}, the power function can be further  bounded in terms of the fill distance 
\[
h_{\Xi_N,\Omega}\triangleq \sup_{x\in \Omega} \inf_{\xi_i\in \Xi_N} \|x-\xi_i\|_\XX. 
\]
In this case, which still covers an enormous class of choices for the vRKHS $\vK$, the controller performance is even more concise, 
\begin{align}
    \limsup_{t\to \infty} \|x(t)-x_r(t)\|_\XX &\lesssim h^s_{\Xi_N,\Omega} \|(I-\Pi_N)f\|_\vK, \\
    &\lesssim h^s_{\Xi_N,\Omega} \|f\|_\vK,  \label{eq:bestbounds} 
\end{align}
for a ``smoothness index'' $s>0$ that depends on the choice of kernel. In such a case, we obtain the two robustness guarantees for  the uncertainty classes and $\calC_{R,\epsilon,N}$ and $\calC_{R}$, respectively,  
\begin{align}
\limsup_{t\to \infty} \|x(t)-x_r(t)\|_\XX &\lesssim h^s_{\Xi_N,\Omega} R \quad \quad \text{ for all systems in Eq. \ref{eq:model1} s.t.  } f\in \calC_R, \label{eq:robustclassbound1} \\
\limsup_{t\to \infty} \|x(t)-x_r(t)\|_\XX     &\lesssim  
 \epsilon \cdot h^s_{\Xi_N,\Omega} \quad \quad \text{for all systems in Eq. \ref{eq:model1} s.t. } f\in \calC_{R,\epsilon,N}.  \label{eq:robustclassbound2}
\end{align}

This overall deterministic analysis has several advantages. (1) The form of the performance bounds in Equation \ref{eq:bestbounds} is concise, and it is very easy to state or interpret geometrically. In fact the authors would argue it is much simpler to state than any of the performance bounds based on concentration inequalities and GP techniques in the dozens of papers cited above on adaptive control methods based on GP methods. An expanded discussion of this comparison of performance guarantees can be found in \cite{kurdilaBook}.  (2) The bounds in Equation \ref{eq:bestbounds} can be applied for any set of centers $\Xi_N$, no matter if they come from an IID random walk that samples a Gaussian process or deterministic measurement process. Of course, for a IID random walk the fill distance $h_{\Xi_N,\Omega}$ is not certain to be small for any sample size $N$.  The theoretical guarantees for methods based on concentration bounds and properties of GPs are based on \textit{retaining all the samples of the random walk}, which for any finite $N$ need not generate a specific instance of a  sample set $\Xi_N$ for which $h_{\Xi_N,\Omega}$ is small. (3) Even if we insist on using the  stochastic analysis associated with IID sampling of a GP, the bounds above can always be used for a specific instance of samples $\Xi_N$ obtained by the IID  measurements. The cited bounds provide a technique whereby we can reject centers given by the random walk  that are too close or generate bases nearly redundant'' with previous samples. In practice we can just stop the ``IID sample and  rejection'' scheme when the after-sampling-computed fill distance is small enough. Again, this will only give performance guarantees for the specific instance of resultant centers. In contrast, the existing GP-based adaptive control above use concentration inequalities that express the confidence and accuracy \textit{for any or all possible IID sample sets $\Xi_N$ where we do not use rejection and keep all centers.} Note that developing extensions of standard GP methods to include mechanisms of outlier rejection, and  deriving rigorous approximation bounds for such modifications of the baseline IID strategy, has been a topic of much concern over the years in the literature on GP approaches. See the efforts in \cite{csato2002sparse,titsias2009variational,bui2017streaming,uhrenholt2021probabilistic} for representative examples.

\section{Background}
\subsection{Symbols and Notation}
In this paper we denote by $\RR$ and $\RR^+$ the real numbers and nonnegative real numbers, respectively. The expression $a \lesssim b$ is used if there is a constant $c>0$ for which  $a \leq c b$, where $c$ is independent of any parameters on which $a$ and $b$ may depend.  This can be convenient to suppress extra  constants which do not play a significant role in subsequent analysis. We write $a\sim b$ if we have $a\lesssim b$ and $a\gtrsim b$.  For any two sets $X,Y$, the vector space of functions from $X \to Y$ is defined as  
  \[
  \calF(X,Y)\triangleq Y^X \triangleq \left \{ f:X \to Y \right \},  
  \]
  with the usual (pointwise) definitions of the vector space operations. 
 We write $\calL(U,V)$ for the collection of all bounded linear operators acting between normed vector spaces $U,V$. In this paper we  often exploit the  basic fact that a linear operator is bounded if and only if it is continuous.  For an operator $T:U\to V$, $\Range{T}$ and $\Null{T}$ are the range and nullspace, respectively, of $T$. We say that $U$ is continuously embedded in $V$ if $U\subset V$ and the canonical embedding $\mathcal{I}:u\in U \mapsto \mathcal{I}(u)=u\in V$ is is a bounded linear operator. In this case we write $U\overset{\mathcal{I}}{\hookrightarrow} V$, and we know that $\|Tu\|_V\leq \|\mathcal{I}\| \|u\|_U$ for all $u\in U$.  

 We denote by $\|\cdot\|_p$ for $1\leq p \leq \infty$ that standard $p$-norm on $\ell^p\triangleq (\RR^n,\|\cdot\|_p)$ given by 
 \begin{align*}
     \|x\|_p \triangleq \left \{ \begin{array}{lll}
     \left (\sum_{i=1}^n |x_i|^p\right)^{1/p} & & \text{ for } 1\leq p <\infty, \\
     \max_{i=1,\ldots,n}|x_i| & & \text{ for } p=\infty.
     \end{array}
     \right .
 \end{align*}
 When a matrix is viewed as an operator $A\in \calL((\RR^n,\|\cdot\|_p),(\RR^m,\|\cdot\|_q))$, the matrix norm $\|\cdot\|_{p,q}$ is defined using the usual operator norm 
 \begin{align*}
     \|A\|_{p,q}\triangleq \sup_{x\not = 0} \frac{\|Ax\|_q}{\|x\|_p}
 \end{align*}
 for $1\leq p,q \leq \infty$. 

 In this paper we make systematic use of RKHS that contain real-valued functions, as well as RKHS that contain real vector-valued functions, over some set $\Omega$. We refer to these as scalar-valued and vector-valued RKHS, respectively, and also denote the latter as vRKHS. 

Since there are many distinct choices of RKHS and vRKHS, we adopt some overall conventions to help the readability of the paper. We denote Euclidean spaces of various dimensions using fonts such as  $\XX,\UU,\YY$, which refer to the state space, space of control values and outputs, respectively.  Using these spaces,  we represent  vRKHS  that contain functions defined on $\Omega$  such as  
\begin{align*}
    \vK&\triangleq \vK(\Omega,\XX), \quad h\in \vK \Rightarrow h:\Omega \to \YY. 
\end{align*}
The reproducing kernels for $\vK$ is written, respectively, in terms of fonts 
\begin{align*}
\Knl&\triangleq \Knl(\cdot,\cdot): \Omega \times \Omega \to \calL(\YY), 
\end{align*}

When $\YY\triangleq \RR$, we use different fonts to distinguish the reproducing kernel and associated native space that contains real-valued functions.  We  denote a scalar-valued kernel  using the font
 $\knl:\XX \times \XX \to \RR$. We write $\sK\triangleq \sK(\XX,\RR)$ for the native space of real-valued functions determined by the scalar-valued kernel $\knl$. 

 \subsection{Point Sets and Distances}
\label{sec:point_sets_distances}
In a number of places in this paper we  use the definition of the  fill distance and minimal separation radius of a finite set of $N$ points $\Xi_N\subset \Omega$. In our applications the set $\Omega$ can be either a bounded subset of a Euclidean space or of a compact manifold. So we initially define these concepts for $(\Omega,d_\Omega)$ a metric space, and subsequently comment on how they apply for certain bounded subsets of Euclidean space or compact manifolds. When the finite set $\Xi_N\subset \Omega$, we  denote by $h_{\Xi_N,\Omega}$ the fill distance of the set $\Xi_N$ in $\Omega$, which is  
\[
h_{\Xi_N,\Omega}\triangleq \sup_{\omega \in \Omega} \min_{\xi_i\in \Xi_N} d_\Omega(\omega,\xi_i). 
\]
The minimal separation distance $r_{\Xi_N}$ is 
\[
r_{\Xi_N}\triangleq \frac{1}{2}\max_{\xi_i,\xi_j\in \Xi_N,\xi_i\not=\xi_j } d_\Omega(\xi_i,\xi_j).
\]
The minimum separation radius can be understood intuitively as the radius of the largest open ball that can be placed between the centers in $\Xi_N$. 
It is immediate from the definition that $r_{\Xi_N}\leq h_{\Xi_N,\Omega}$. But it is always possible to define families of centers $\Xi_N$ for which $r_{\Xi_N}\to 0$ as $N\to \infty$, but for which $h_{\Xi_N,\Omega}$ is bounded away from zero by a constant that is independent of $N$. In this sense the fill distance is a better indicator of the unformity of the centers in $\Xi_N$ in the set $\Omega$. We say that family of centers $\Xi_N$ is quasiuniform in $\Omega$ if there is a  constant $C>0$ such that 
\[
 r_{\Xi_N} \leq h_{\Xi_N,\Omega} \leq C r_{\Xi_N}. 
\]
It is often assumed that  samples  are quasiuniform in a metric space $(\Omega,d_\Omega)$ since they then obey scaling laws that agree with our intuitions that hold for  uniformly defined samples in a parallelopiped in $\RR^n$. 

For instance, the number $N$ of uniform samples in a parallelopiped like $\Omega\triangleq [0,1]^n$ scale like 
\[
N=N(\Xi_N,\Omega)\sim \frac{1}{h^{n}_{\Xi_N,\Omega}} \sim \frac{1}{r_{\Xi_N}^n}. 
\]
This is easy to argue since the volume of $\Omega$ is one, and 
\[
\mu(\Omega)=1\sim N\cdot(h_{\Xi_N,\Omega}^n)
\]
where $\mu$ is Lebesgue measure on $\RR^n$ and $h^n_{\Xi_N,\Omega}$ is the volume of a cube having side length $h_{\Xi_N,\Omega}$. The same scaling holds more generally for quasiuniform samples in any parallopiped  that is a subset of $\RR^n$. 

These scalings also hold for $\ell$-dimensional, compact, smooth, Riemannian  manifolds $\calM$ that are regularly embedded in $\RR^n$, but with the dimension $n$ in the estimate above replaced by  the dimension $\ell$ of the manifold.  For quasiuniform samples in $\calM\subset \RR^n$, we have 
\[
N=N(\Xi_N,\calM)\sim \frac{1}{h^{\ell}_{\Xi_N,\calM}}.
\]
This can be proven using the following volume comparison \cite{narcowich2012,hangelbroek2018inverse} for compact Riemannian manifolds. For any such smooth, compact, $\ell$-dimensional Riemannian manifold $\calM$, there are constants $\alpha,\beta>0$ such that 
\[
\alpha r^\ell \leq \mu(B_{r}(\xi)) \leq \beta r^{\ell}  \quad \quad \text{ for all } \xi\in \calM, 
\]
where $\mu$ is the volume measure on the manifold $\calM$ and $B_r(\xi)$ is the geodesic open ball of radius $r$ centered at $\xi\in \calM$. Here $r>0$ is any radius that is less than or equal to the diameter of the compact manifold $\calM$.  If we take the radius $r$ to be the minimal separation radius $r_{\Xi_N}$, the above inequalities imply that 
\begin{align*}
\sum_{i=1}^N \mu(B_{r_{\Xi_N}}(\xi_i))&\leq \mu(\calM),\\
N\cdot(\alpha r^\ell_{\Xi_N})  &\leq \mu(M), 
\end{align*}
since this collection of balls is disjoint and contained in $\calM$. Hence there is a constant such that $N\lesssim r^{-\ell}_{\Xi_N}$. On the other hand, we know that 
\begin{align*}
    \calM & =\bigcup_{1=1}^N B_{h_{\Xi_N,\calM}}(\xi_i), \\
    \mu(M) &\leq N\mu(B_{h_{\Xi_N,\calM}}(\xi_i)) \leq N\beta h_{\Xi_N,\calM}^\ell, 
\end{align*}
so $N\gtrsim h^{-\ell}_{\Xi_N,\calM}$. From the equivalence of the fill distance and minimum separation for quasiuniform samples, we obtain $N\sim r^{-\ell}_{\Xi_N}\sim h^{-\ell}_{\Xi_N,\Omega}$.

\subsection{Operator-Valued Kernels and vRKHS}
\label{sec:operatorvRKHS}

Suppose that we are interested in using reproducing kernel techniques to construct functions $f:X \to Y$ over the set  $X$ that take values in the separable Hilbert space  $Y$. 
An admissible operator-valued kernel  is a mapping $\Knl:X \times X \to \calL(Y)$
that satisfies 
\begin{enumerate}
    \item the symmetry property 
    $$ \Knl(x_1,x_2)=\Knl(x_2,x_1)^*, \quad \quad \text{ for all } x_1,x_2 \in X, 
    $$
    \item and the non-negativity property 
    \begin{align}
    \sum_{i=1}^N \sum_{j=1}^N\alpha_i\alpha_j\left (\Knl(\xi_i,\xi_j)y_i,y_j \right)_Y \geq 0    \label{eq:nonnegK}
    \end{align}
    for all collections of $N$ coefficients $\{\alpha_1,\ldots,\alpha_N\}\subset \RR$, selection of centers $ \Xi_N\triangleq \{\xi_1,\ldots,\xi_N\}$ contained in $X$, and directions $\{y_1,\ldots,y_N\}\subset Y$. 
\end{enumerate}
 When $\Knl$ is such an operator-valued kernel, the corresponding vRKHS is precisely 
\begin{align*}
    \vK&\triangleq \vK(X,Y)=\overline{\text{span} \left \{ \Knl_x y \ | \ x\in X, y\in Y\right \}},
\end{align*}
where we define $\Knl_x(\cdot)\triangleq \Knl(\cdot,x)$ for any $x\in X$. The closure above means that for any $f\in \vK$ there is a sequence of coefficients $\alpha_{N,i}\subset \RR$, centers $\xi_{N,i}\subset X$, and directions $y_{N,i}\subset Y$ such that 
\begin{align*}
    \lim_{N\to \infty} \|f- \sum_{i=1}^N \alpha_{N,i}\Knl_{\xi_{N,i}}y_{N,i}\|_\vK =0. 
\end{align*}
See \cite{carmeli10,micchelli2005learning,paulsen} for a detailed description of this standard construction where the limits of finite dimensional approximations as above are  determined from  the concrete expressions as they appear in Equation \ref{eq:nonnegK}. 

 Instead of the positivity condition above, it is common that positivity is described in terms of the generalized Grammian matrix associated with a set of centers $\Xi_N\subset X$. The generalized Grammian matrix is given by 
\begin{align*}
\KK_N\triangleq \left [\begin{matrix}
    \Knl(\xi_1,\xi_1) & \cdots & \Knl(\xi_1,\xi_N)\\
    \vdots & \ddots & \vdots \\
    \Knl(\xi_N,\xi_1)& \cdots & \Knl(\xi_N,\xi_N)
\end{matrix}\right ]\in \calL(Y^N).
\end{align*}
The operator-valued kernel is then of positive type if the generalized Grammian matrix $\KK_N$ is positive semidefinite for any choice of centers $\Xi_N\subset X$. The operator-valued kernel  is of strictly positive type, or positive definite,  if $\KK_N$ is positive definite for any collection of $N$ distinct centers $\Xi_N\subset X$.

The definition of an admissible operator-valued kernel above is very general and does not impose any continuity properties on the kernel $\Knl$, nor on the vRKHS $\vK$ generated by $\Knl$. 
When $X$ is a topological space, we  say the operator-valued kernel $\Knl$ is a Mercer kernel if the vRKHS $\vK\triangleq \vK(X,Y)$ defined by $\Knl$ is a subspace of $C(X,Y)$, where the latter is endowed with the compact-open topology. \cite{carmeli10}. We only consider cases where the topology on $X$ is induced by a metric, in which case the compact-open topology is precisely the topology of uniform convergence over compact subsets of $X$.

The vRKHS defined in this way have a number of extraordinary properties that make them particularly effective in posing and solving many classical problems of regression or machine learning theory. One foundational property is that the evaluation operator $E_x$ at a location $x\in X$, which  is defined by the identity $E_x f\triangleq f(x)\in \YY$ for all $x\in X$, is a bounded linear operator $E_x\in \calL(\vK,Y)$ satisfying  the \textit{reproducing formula}
\[
\left (\Knl_x y,f \right)_\vK = \left (y,E_x f \right)_Y = (y,f(x))_Y \quad \quad \text{ for all } f\in \vK, y\in Y.
\]
That is, $E_x^*=(E_x)^*=\Knl_x$ for all $x\in X$. This identity guarantees, among other uses, that $\Knl_x\in \calL(Y,\vK)$ for all $x\in X$. 
Among other things, this inequality implies that  the norm of $f$ in the vRKHS $\vK$ dominates the pointwise norm in $Y$ of its value $f(x)\in Y$, in the sense that 
\begin{align}
f(x)&\leq \sqrt{\|\Knl(x,x)\|_{\calL(Y)}}\|f\|_\vK \quad \quad \text{ for all } x\in X, f\in \vK. \label{eq:K_dominates_pointwise}
\end{align}

This inequality is important in applications since it guarantees that there are rather simple conditions to ensure that an operator-valued kernel is a Mercer kernel.  Proposition 2 of \cite{carmeli10} uses this property to establish that an operator-valued kernel $\Knl:X\times X \to \calL(Y)$ is a Mercer kernel if and only if 
\begin{enumerate}
    \item $\Knl_xy\in C(X,Y)$ for all $x\in X$, $y\in Y$, and 
    \item the mapping $x\mapsto \|\Knl(x,x)\|_{\calL(Y)}$ is locally bounded for each $x\in X$. 
\end{enumerate}
To simplify the theory in this paper, we assume  that the operator-valued kernel $\Knl$ is bounded on the diagonal, which is not too restrictive in our applications to adaptive control. We say that $\Knl$ is bounded on the diagonal if  there is a constant $\bar{\knl}>0$ such that 
\begin{align*}
    \|\Knl(x,x)\|_{\calL(Y)}\leq \bar{\knl}^2 \quad \text{ for all } x\in X.  
\end{align*}
When this property holds, we have the uniform bound $\|E_x\|=\|E_x^*\|=\|\Knl_x\|\leq \bar{\knl}$ for all $x\in X$.   This follows from the inequalities 
\begin{align*}
    \|E^*_xy\|^2_\vK &= \left (E_xE_x^* y,y \right )_Y \\ &=\left(\Knl(x,x)y,y\right)_Y 
    \leq \bar{\knl}^2\|y\|_\YY^2 \quad \quad \text{ for all } x\in X,y\in Y.
\end{align*}
Thie property proves   essential in our later study of the Lyapunov stability and convergence of adaptive control schemes.
If the operator-valued kernel $\Knl$ is bounded on the diagonal and the mapping $x\mapsto \|\Knl(x,x)\|$ is continuous, it  then easily follows from Equation \ref{eq:K_dominates_pointwise}, as well as (1) and (2) above, that 
\[
\vK \hookrightarrow C_b(X,Y).
\]
We only consider operator-valued kernels in this paper that are continuously embedded in $C_b(X,Y)$, as above. 

The appendix contains two technical theorems that are used repeatedly in this paper. These include Theorem \ref{th:vector_feature} that characterizes feature operators and feature spaces, and Theorem \ref{th:vectorZSHS} that applies this theorem to construct vRKHSs generated by a subset or vRKHSs that contain restrictions to a subset. Suppose that $\Omega\subseteq \XX$ and the operator-valued kernel $\Knl:\XX \times \XX \to \calL(\YY)$ generates the vRKHS $\vK\triangleq \vK(\XX,\YY)$. We define the vRKHS   $\vK_\Omega$ and $\vR_\Omega$ to be given by  
\begin{align*}
    \vK_\Omega &\triangleq \vK_\Omega(\XX,\YY) = \overline{\text{span}\left \{ \Knl_xy \ | \ x\in \Omega, y\in \YY\right \}}\subseteq \vK, \\
    \vR_\Omega & \triangleq \vR_\Omega(\Omega,\YY) = T_\Omega(\vK)= \left \{ r:\Omega \to \YY \ | \ r=T_\Omega g \triangleq g|_{\Omega}, g\in \vK \right \}, 
\end{align*}
where $T_\Omega(\cdot)\triangleq (\cdot)|_\Omega$ is the trace or restriction operator. 
We refer to $\vK_\Omega$ as the vRKHS generated by the set $\Omega \subseteq \XX$, while $\vR_\Omega$ is just referred to as the space of restrictions of functions in $\vK$. Often in this paper we say that functions in $\vK$ or $\vK_\Omega\subseteq \vK$ are global functions since they are defined on all of $\XX$, while we say functions in $\vR_\Omega$ are local functions since they are defined only over $\Omega$. The relationships between the operator-valued kernels that defined $\vK$, $\vK_\Omega$, and $\vR_\Omega$ are discussed in detail in the appendix. 

Here we only note that the operator-valued kernels for $\vK$ and $\vR_\Omega$ are defined in such a way that $T_\Omega\in \calL(\vK,\vR_\Omega)$, and therefore 
\[
T_\Omega^*\in \calL(\vR_\Omega,\vK).
\]
In fact, the adjoint operator $T_\Omega^*$, since it maps a local function in $\vR_\Omega$  to a global function in $\vK$, defines a canonical extension operator 
\[
\Ext_\Omega\triangleq T_\Omega^* \in \calL(\vR_\Omega,\vK).
\]
We use the extension operator $\Ext_\Omega$ and restriction operator $T_\Omega$ to pass back and forth between the global vRKHSs $\vK$ or $\vK_\Omega$  and the  local vRKHS $\vR_\Omega$. See the appendix for the details on the properties and relationships of these operators and spaces.

\begin{comment}
The above considerations make it rather straightforward to generate vRKHS that are continuously embedded in the well-known  space $C_b(X,Y)$.  Here we close this short introduction by recalling a way of ensuring that one vRKHS is continuously embedded in another.    Suppose that $\Knl_1$ and $\Knl_2$ are two admissible operator kernels that define the vRKHS $(\vK_1,\|\cdot\|_{\vK_1})$ and $(\vK_2,\|\cdot\|_{\vK_2})$, respectively. The following simple, yet powerful, theorem gives an easily verified  condition to conclude that one of  these spaces is continuously embedded in the other.
\begin{theorem}[Theorem 6.25, \cite{paulsen}]
  Suppose that $\Knl_1$ and $\Knl_2$ are two admissible operator kernels that define the vRKHS $(\vH_1,\|\cdot\|_{\vH_1})$ and $(\vH_2,\|\cdot\|_{\vH_2})$, respectively. If there is a constant $c>0$ such that 
  \begin{align}
  \Knl_1(x_1,x_2)\leq c^2 \Knl_2(x_1,x_2) \quad \quad \text{ for all } x_1,x_2\in X, \label{eq:K1LEQK2}
  \end{align}
  then we have the continuous inclusion 
  \begin{align}
  \vK_1\overset{\Inj}{\hookrightarrow} \vK_2 
  \label{eq:H1hookH2}
  \end{align}
  where $\Inj:\vK_1 \to \vK_2$ is the canonical injection 
  and $\|\Inj\|\leq c$. That is, we have 
  \[
  \|g\|_{\vK_2}\leq \|\Inj\|\|g\|_{\vK_1} \leq c\|g\|_{\vK_1} \quad \quad \text{ for all } g\in \vK_1. 
  \]
 On the other hand, if Equation \ref{eq:H1hookH2} holds, then Equation \ref{eq:K1LEQK2} holds with $c=\|\Inj\|$.  
\end{theorem}
\end{comment}

\subsection{Orthogonal Projection,  Restrictions, and Extensions}
\label{sec:interpolation}
This paper also  makes systematic use of interpolation and projection over subspaces $\vK_N\subseteq \vK$  of the vRKHS $\vK\in \vK(\XX,\YY)$  generated by an operator-valued kernel $\Knl(x_1,x_2)\in \calL(\YY)$. In this section we summarize a special case of the overall general  approach presented in \cite{wittwar2018interpolation,wittwar2022approximation} that analyzes error of approximation in terms of  power functions for vRKHSs defined in terms of operator-valued kernels. Throughout this paper we assume that the operator-valued kernel is strictly positive, which simplifies several proofs and is typical for our applications in adaptive control. See \cite{wittwar2018interpolation,wittwar2022approximation} for the theory when the   operator-valued kernels that are only positive semidefinite.  The analysis for strictly positive operator-valued kernels in this section studies the relationship between approximations generated by $\vK$-orthogonal, $\vK_\Omega$-orthogonal, and $\vR_\Omega$-orthogonal projections in terms of the above canonical extension operators $\Ext_\Omega=T^*_\Omega$ and restriction/trace operator $T_\Omega$.

In the paper we assume that the operator-valued kernel is strictly positive.  We always end up defining approximations in $\vK$ in terms of finite dimensional subspaces 
\begin{align*}
\vK_N  &\triangleq \text{span} \left \{ \Knl_{\xi_i}e_j |  \xi_i\in \Xi_N, 1\leq i \leq N, 1\leq j \leq m,\right \}\\
&\triangleq \text{span} \left \{ \Knl_{\xi_i}y |  \xi_i\in \Xi_N, 1\leq i \leq N, y\in \YY\right \}.
\end{align*}
where $e_j$ is the $j^{th}$ canonical basis vector in $\YY\triangleq \RR^m$. The strict positivity of the operator kernel implies that we always have 
\[
\text{dim}(\vK_N)=mN. 
\]

We say that an approximation $\hat{f}_N\in \vK_N$ interpolates the function $f\in \vK$ at the centers $\Xi_N\subset \XX$ if we have 
\[
\hat{f}_N(\xi_i)=f(\xi_i) \quad \quad \text{ for } 1\leq i\leq N.  
\]
For a given the set of centers $\Xi$, and assuming as above that $\Knl$ is strictly positive,  we denote by $\mathcal{I}_N:\vK\to \vK_N$ the uniquely defined interpolation operator that generates the  approximation $\hat{f}_N=\mathcal{I}_N f$  that interpolates $f$. 
It is immediate the $\vK$-orthogonal projection operator $\bm{\Pi}_N:\vK\to \vK_N$ is identical to the interpolation operator. By definition of the  $\vK$-orthogonal projection operator on $\vK_N$, we have 
\begin{align*}
    \left (\bm{\Pi}_Nf-f,\Knl_{\xi_i}y  \right)_\vK = 0 \quad \quad \text{ for all  } \xi_i\in \Xi_N, y\in \YY.
\end{align*}
But since $\Knl_{\xi_i}^*=E_{\xi_i}$, by the reproducing property in the vRKHS $\vK$, we know that 
\begin{align*}
\left (E_{\xi_i}(\bm{\Pi}_Nf-f),y  \right)_\YY &= 0,  \\
\left ((\bm{\Pi}_Nf)(\xi_i)-f(\xi_i)),y  \right)_\YY &= 0 \quad \quad \text{ for all } \xi_i\in \Xi_N, y\in \YY. 
\end{align*}
Thus, $\bm{\Pi}_Nf$ interpolates the function $f$ at the $N$ centers $\Xi_N$, and $\bm{\Pi}_N=\mathcal{I}_N$.  

When we define the basis vectors $\phi_{ij}\triangleq \Knl_{\xi_i}e_j$, a typical function in $\vK_N$ can be written
\begin{align*}
    f=\sum_{i=1}^N \sum_{j=1}^m\theta_{ij}\phi_{ij}=\sum_{i=1}^N \vartheta_i^T \bm{\varphi}_i =\Theta_N^T \Phi_N,
\end{align*}
where $\vartheta_i\in \RR^{m\times 1}$, $\bm{\varphi}_i(x)\in \RR^{m\times 1}$, $\Theta_N=\{\vartheta_1^T,\ldots,\vartheta_N^T\}^T\in \RR^{mN\times 1}$, $\bm{\varphi}_i(x)\triangleq\{\phi_{i1}(x),\ldots,\phi_{im}(x) \}^T\in \RR^{m\times 1}$, and $\Phi_N(x)=\{\bm{\varphi}_1(x)^T,\cdots, \bm{\varphi}_N(x)^T\}^T\in \RR^{mN\times 1}$. It is easy to show that, relative to these choices and organization of bases, the interpolation or $\vK$-orthogonal projection operator has the coordinate representation $\bm{\Pi}_Nf=\Theta_N^T\Phi_N$ where the coefficients are obtained by solving the linear system 
\begin{align}
\KK_N \Theta_N = F_N \label{eq:interpolation_equation}
\end{align}
where $F_N\triangleq \{f(\xi_1)^T,\cdots, f(\xi_N)\}^T\in \RR^{mN\times 1}$ and $\KK_N$ is the generalized Grammian matrix of $\vK_N$ in $\vK$. 

We can view the subspace $\vK_N$ as the subspace that is generated by the subset $\Xi_N\subset \XX$, in the sense described in Theorem \ref{th:vectorZSHS}, and therefore it is a vRKHS for the operator-valued kernel 
\[
\Knl_N(x_1,x_2)\triangleq E_{x_1}\bm{\Pi}_{N}E^*_{x_2} \quad \quad \text{ for all } x_1,x_2 \in \XX,  
\]
where $\bm{\Pi}_{N}$ is the $\vK$-orthogonal projection onto $\vK_N$. See Equation  \ref{eq:operator_kernel_U} above. By substituting the results from Equation \ref{eq:interpolation_equation} in this definition, we find the representation 
\begin{align*}
    \Knl_N(x_1,x_2)&=\begin{bmatrix}\Knl(x_1,\xi_1)&\Knl(x_1,\xi_2)&\cdots&\Knl(x_1,\xi_N)\end{bmatrix}\KK_N^{-1}\begin{bmatrix}\Knl(\xi_1,x_2)\\ \Knl(\xi_2,x_2)\\ \vdots\\  \Knl(\xi_N,x_2) \end{bmatrix},\\
    &\triangleq\Knl(x_1,\Xi_N)\KK^{-1}_N\Knl(x_2,\Xi_N)^*,
\end{align*}
where $\Knl(x_1,\Xi_N)\in \RR^{m\times (mN)}$ and $\Knl(x_2,\Xi_N)^*\in \RR^{(mN)\times m}$. See \cite{wittwar2018interpolation,wittwar2022approximation} for alternative derivations that illustrate this fact, including the more general situation when the operator-valued kernel is only positive semidefinite.  

The above organization and explanation of interpolation and orthogonal projections in $\vK$ is actually generic, and  it can be applied to any vRKHS generated by a strictly positive operator kernel. Now let $\vR_\Omega$ be the space of restrictions of functions in $\vK$ to the set $\Omega$, where we equip $\vR_\Omega$ with the restricted kernel described as in the last section, 
\[
\Rnl_\Omega(\omega_1,\omega_2)\triangleq \Knl(\omega_1,\omega_2)\in \calL(\YY)  \quad \quad \text{ for all } \omega_1,\omega_2 \in \Omega. 
\]
It is immediate that the strict positivity of $\Knl$ implies that the kernel $\Rnl_\Omega$ is also of strictly positive type,  since the generalized Grammians $\KK_N=\RR_N$ for centers $\Xi_N\subset \Omega$. 
Similarly to the above, we define the finite dimensional spaces 
\[
\vR_{N}  \triangleq \text{span} \left \{ \Rnl_{\Omega,\xi_i}e_j | 1\leq j \leq m, \xi_i\in \Xi_N, 1\leq i \leq N\right \}
\]
where the set of centers $\Xi_N\subset \Omega\subset \XX$. Note that the basis $\psi_{ij}=\Rnl_{\Omega,\xi_i}e_j$ here is just the  restriction of the basis $\phi_{ij}\triangleq \Knl_{\xi_i}e_j$ used above for approximations of functions  over $\XX$, satisfying  $\phi_{ij}=\Ext_\Omega(\psi_{ij})$ and $\psi_{ij}=T_\Omega(\phi_{ij})$.   When we define the coordinate representation of the $\vR_\Omega$-orthogonal projection $\tilde{\bm{\Pi}}_{N}:\vR_\Omega \to \vR_{N}$ as $\tilde{\Pi}_Nr=\tilde{\Theta}_N^T\Psi_N$ for any function $r\in \vR_\Omega$, we find  the coefficients match those above, with $\tilde{\Theta}_N\equiv \Theta_N$. This is implied by the fact that, as long as the centers $\Xi_N\subset \Omega$, the generalized Grammians of $\vK_N$ in $\vK$ and of $\vR_N$ in $\vR_\Omega$ are identical, that is, $\KK_N=\RR_N$.  This line of reasoning shows that,  whenever the centers $\Xi_N\subset \Omega$, we have the identities  
\begin{align*}
    \bm{\Pi}_N \Ext_\Omega &= \Ext_\Omega \tilde{\bm{\Pi}}_N,\\
    T_\Omega \bm{\Pi}_N & = \tilde{\bm{\Pi}}_NT_\Omega. 
\end{align*}
%
%% Power functions
%
\subsection{Power Functions and Orthogonal Projections for vRKHSs}
It is now standard in discussions  of approximations in RKHS $\sH$ that contain real-valued functions to use the power function to assess their accuracy. See \cite{wendland,schaback94,kurdilaBook} for a detailed account of the theory for scalar-valued RKHS. In this article we frame the overall problem via a generalization of the power function that is suitable for the study of vRKHS.  The overall approach is based on the groundwork developed in \cite{wittwar2018interpolation,wittwar2022approximation}.

Let $\vK\triangleq \vK(\XX,\YY)$ be a vRKHS that is defined in terms of a strictly positive operator kernel $\Knl(x_1,x_2)\in \calL(\YY)$ for all $x_1,x_2\in \XX$. When $\vU$ is a closed subspace of $\vK$ and $\Pi_\vU:\vK\to \vU$ is the $\vH$-orthogonal projection onto $\vU$,   the generalized power function $\Pwr_{\vU}\triangleq \Pwr_{\vU,\vK}: \vK^*\to \RR^+$ is defined by the identity
\begin{align*}
    \Pwr_{\vU}(h^*)&\triangleq \sup_{f\in \vK \backslash 0} \frac{h^*\left ( (I-\Pi_\vU)f\right )}{\|f\|_\vK} \quad \quad \text{ for all } h^*\in \vK^*,
\end{align*}
where $\vK^*$ is the (topological) dual space of the Hilbert space $\vK$. When we denote by $h\triangleq h(h^*)\in \vK$ the unique Riesz representer of $h^*\in \vK^*$, this definition can alternatively be written (see Corollary 2.9 in \cite{wittwar2018interpolation}) as 
\[
\Pwr_\vU(h^*)\triangleq \|(I-\Pi_\vU)h\|_{\vH}=\|\Pi_{\vU^\perp}h\|_\vH \quad \quad \text{  for all } h^* \in \vK^*.
\]

This identity defines the action of the power function for any functional $h^*\in \vK^*$, but most often applications of the definition are used for the functional $h^*_{x,y}(f)=\left(\Knl_xy,f \right)_{\vH}=(y,f(x))_\YY$ associated with the fixed center $x\in \XX$ and output (direction) $y\in \YY$. In this case References \cite{wittwar2018interpolation,wittwar2022approximation}  introduce the power function for vRKHS that has two arguments. We use  the notation 
\[
\Pwr_\vU(x,y)\triangleq \Pwr_\vU\left (h^*_{x,y} \right) \quad \quad \text{ for all } x\in \XX, y\in \YY, 
\]
and refer to $\Pwr_\vU(x,y)\triangleq \Pwr_{\vU,\vK}$ as the power function of the closed subspace $\vU$ of  $\vK$ at the center $x\in \XX$ in the output direction $y\in \YY$. For brevity we just refer to $\Pwr_\vU$ as the power function of the closed subspace $\vU$ when the definition of the container space $\vK$ is clear.  However, in some instances, we must use the more explicit notation $\Pwr_{\vU,\vK}$ to be unambiguous. 

We conclude this section by summarizing an amalgam of some the important common properties of the generalized power function $\Pwr_\vU(x,y)$ derived at various points in Section 2 of \cite{wittwar2018interpolation}. 
\begin{theorem}
Under the operating assumptions above, the following hold: 
\begin{enumerate}
    \item For any center $x\in \XX$ and output direction $y\in \YY,$
    \begin{align*}
        \Pwr_{\vU}(x,y)= \left ((\Knl(x,x)-\Knl_\vU(x,x))y,y \right)_\YY = \left( \Knl_{\vU^\perp}(x,x)y,y\right)_\YY.
    \end{align*}
    \item For any center $x\in \XX$ and output direction $y\in \YY$, we have the following pointwise error bounds 
    \begin{align*}
(E_x(I-\Pi_\vU)f,y)_\YY &\leq \Pwr_\vU(x,y) \|(I-\Pi_\vU))f\|_\vK \leq \Pwr_\vU(x,y)\|f\|_\vK,\\
\|E_x(I-\Pi_\vU)f\|_{2} &\leq \sqrt{\|\Knl(x,x)-\Knl_{\vU}(x,x)\|_{2,2}}\|(I-\Pi_\vU))f\|_\vK, \\
\|E_x(I-\Pi_\vU)f\|_{\infty},&\leq \max_{i=1,\ldots,m}\sqrt{|\Knl_{ii}(x,x)-\Knl_{\vU,ii}(x,x)|}\|(I-\Pi_\vU)f\|_\vK, \\
\|E_x(I-\Pi_\vU)f\|_{1} &\leq  \sqrt{m}\sqrt{\|\Knl(x,x)-\Knl_{\vU}(x,x)\|_2} \|(I-\Pi_\vU))f\|_\vK.
    \end{align*}
\end{enumerate}
\end{theorem}

For brevity, we introduce the notations
\begin{align}
    \bar{\Pwr}_{2,\vU}(x)&\triangleq \sqrt{\|\Knl(x,x)-\Knl_{\vU}(x,x)\|_{2,2}}, \label{eq:PWR_2}\\
    \bar{\Pwr}_{\infty,\vU}(x)&\triangleq \max_{i=1,\ldots,m}\sqrt{|\Knl_{ii}(x,x)-\Knl_{\vU,ii}(x,x)|} \label{eq:PWR_infty}
\end{align}
to refer to the variants of  $\Pwr_{\vU,\vK}$ above, which  simplifies the form of equations in the remainder of the paper. 

Finally, while we have defined the power function above for a general closed subspace $\vU\subset \vK$, it is most commonly the case that we apply these definitions for a finite dimensional subspace $\vK_N\subset \vK$. In this case we just denote the power function of $\vK_N$ in $\vK$ as $\bar{\Pwr}_{\vK_N}$, so that 
\begin{align}
    \bar{\Pwr}_{2,\vK_N}(x)&\triangleq \sqrt{\|\Knl(x,x)-\Knl_{N}(x,x)\|_{2,2}}, \label{eq:PWR_2_N},\\
    \bar{\Pwr}_{\infty,\vK_N}(x)&\triangleq \max_{i=1,\ldots,m}\sqrt{|\Knl_{ii}(x,x)-\Knl_{N,ii}(x,x)|}, \label{eq:PWR_KN_infty_N}
\end{align}
for all $x\in \XX$ and $\Knl_N$ is the operator-valued kernel of $\vK_N\subseteq \vK$. 

One important example where the explicit notation that stipulates  the subspace and container space arises in the discussion of the power function of a subspace $\vK_N\subseteq \vK_\Omega$, where $\vK_\Omega$ is the closed subspace of $\vK$ that is generated by the subset $\Omega\subset \XX$.  The power function of the subspace $\vK_N$ in $\vK_\Omega$ is not the same as the power function of $\vK_N$ in the orginal space $\vK$.  In this case we have 
\begin{align}
    \bar{\Pwr}_{2,\vK_{N},\vK_\Omega}(x)&\triangleq \sqrt{\|\Knl_\Omega(x,x)-\Knl_{{\Omega,N}}(x,x)\|_{2,2}}, \label{eq:PWR_2_omega},\\
    \bar{\Pwr}_{\infty,\vK_{N},\vK_\Omega}(x)&\triangleq \max_{i=1,\ldots,m}\sqrt{|\Knl_{\Omega,{ii}}(x,x)-\Knl_{\Omega,N,ii}(x,x)|}, \label{eq:PWR_Omega_KN_infty_omega}
\end{align}
for all $x\in \XX$, where $\Knl_{\Omega}$ is the operator-valued kernel of $\vK_\Omega$ and $\Knl_{\Omega,N}$ is the operator-valued kernel of $\vK_N\subset \vK_\Omega$.  We try to avoid this ugly  notation wherever possible, but sometimes unfortunately it is unavoidable. 

Entirely analogous definitions of the power function $\bar{\Pwr}_{\vR_N}$ hold for finite dimensional spaces $\vR_N$ contained in the space of restrictions $\vR_\Omega$. In this case we write 
\begin{align}
    \bar{\Pwr}_{2,\vR_N}(\omega)&\triangleq \sqrt{\|\Rnl_\Omega(\omega,\omega)-\Rnl_{\Omega,N}(\omega,\omega)\|_{2,2}}, \label{eq:PWR_RN_2},\\
    \bar{\Pwr}_{\infty,\vR_N}(\omega)&\triangleq \max_{i=1,\ldots,m}\sqrt{|\Rnl_{\Omega,ii}(\omega,\omega)-\Rnl_{\Omega,N,ii}(\omega,\omega)|}, \label{eq:PWR_RN_infty}
\end{align}
for all $\omega\in \Omega$, where $\Rnl_{\Omega,N}$ is the operator-valued kernel of $\vR_N\subseteq \vR_\Omega$.

\section{Maneuver or Trajectory vRKHSs}
\label{sec:maneuver}

In this section we establish a few of the primary contributions of this paper. We define  and study  the approximation properties of the maneuver vRKHSs $\vK_\calM$ that are used to define functional uncertainty classes for nonparametric adaptive control methods. The discussion in this section is careful to emphasize that some of the function spaces below are referred to as  \textit{global} in the sense that they contain functions defined over all of the state space $\XX$. These include the initial vRKHS $\vK=\vK(\XX,\YY)$ and the maneuver vRKHS $\vK_\calM=\vK_\calM(\XX,\YY)$. Other function spaces, in particular $\vR_\calM=\vR_\calM(\calM,\YY)$,  are \textit{local} and only contain functions that are defined on the subset $\calM\subset \XX$.

The uncertainty  $f$  appearing in the governing ODEs must be defined on all of the state space $\XX$ for well-posedness of the ODEs. Hence, we always choose the uncertainty $f$ in the maneuver vRKHS $\vK_\calM\subseteq \vK$, which contains globally defined functions.  On the other hand, the samples $\Xi_N$  used for approximations are always selected in the subset $\Xi_N\subset \calM$, and the local space $\vR_\calM$ is used to study convergence rates of approximations. The primary theorems in this section establish how rates of approximations in the local space $\vR_\calM$ can be used to infer rates of convergence in the maneuver vRKHS $\vK_\calM$ of globally defined functions. 

In preparation for our study of trajectory or maneuver spaces containing functional uncertainties, we recall how the doubling trick described in \cite{wendland, hangelbroek2024extending} can be extended to the setting of vector-valued vRKHS defined in terms of general operator-valued kernels.  We suppose that the operator-valued  kernel $\Knl:\XX \times \XX \to \calL(\YY)=\RR^{m\times m}$ defines the vRKHS $\vK(\XX,\YY)$, and that the kernel $\Knl$ is bounded on the diagonal in the sense that there is a constant $\bar{\knl}>0$ such that 
\[
\|\Knl(x,x)\|\leq \bar{\knl}^2 \quad \quad \text{ for all } x\in \XX.  
\]
As discussed in Section \ref{sec:operatorvRKHS}, this ensures that we have the uniform operator bound $\|E_x\|=\|E^*_x\|=\|\Knl_x\|\leq \bar{\knl}$ for all $x\in \XX$.   

 We also fix  a subset  $\mathcal{M}\subset \XX \triangleq \RR^n$ that is an $\ell$-dimensional,  compact, smooth, connected, $\ell$-dimensional Riemannian submanifold that is regularly embedded in $\XX$, and we use it to  define the space of restrictions $\vR_{\calM}$   and the maneuver vRKHS $\vK_\calM\triangleq \Ext_\calM(\vR_\calM)$ that contains globally defined functions.  It is assumed that $\Rnl_\Omega$ is a Mercer kernel so that Proposition 2 of \cite{carmeli10} shows 
 \[
 \vR_\calM\hookrightarrow C(\calM,\YY). 
 \]
 Note also that Proposition 2 of \cite{carmeli10} ensures that $\Rnl_{\calM}$ is a Mercer kernel if and only if the map $\omega \mapsto \|\Rnl_\Omega(\omega,\omega)\|$ is locally bounded and $\Rnl_{\Omega}(\cdot,\omega)y\in C(\calM,\YY)$ for all  $\omega \in \Omega$ and $y\in \YY$.  The former condition is guaranteed by our assumption that $\Knl$ is bounded on the diagonal, and the latter property is used in our analysis of the integral operator below. 
 
 We also define the  Lebesgue space $L^2_\mu(\mathcal{M},\YY)$ of $\YY$-valued functions over $\mathcal{M}$, which is  endowed with the usual norm
\[
\|r\|_{L^2(\mathcal{M},\YY)}\triangleq \sqrt{\int_\calM \|r(\xi)\|^2_\YY \mu(d\xi)}, 
\]
where $\mu$ is the volume measure on the manifold $\calM$.  We  define the linear integral operator 
\begin{align*}
    (Lv)(\xi)\triangleq \int_\calM \Rnl_\calM(\xi,\eta)v(\eta)\mu(d\eta),  
\end{align*}
 where the integral above is the Bochner integral of the function $\Rnl_\calM(\cdot,\eta)v(\eta)\in \vR_\calM$. 
The integral operator  $L:L^2(\calM,\YY)\to \vR_\calM(\calM,\YY)$ is a bounded linear operator since 
\begin{align*}
\|Lv\|_{\vR_\calM} &\leq \int_\calM \|\Rnl_\calM(\cdot,\xi)v(\xi)\|_{\vR_\calM} \mu(d\xi) \\ &= \int_\calM \sqrt{\left (\Rnl_\calM(\cdot,\xi)v(\xi),\Rnl_\calM(\cdot,\xi)v(\xi) \right)_{\vR_\Omega}} \mu(d\xi)\\
&= \int_\calM \sqrt{\left (\Rnl^*_\calM(\cdot,\xi)\Rnl_\calM(\cdot,\xi)v(\xi),v(\xi) \right)_{\YY}} \mu(d\xi)\\
&=\int_\calM \sqrt{\left (\Rnl_\calM(\xi,\xi)v(\xi),v(\xi) \right)_{\YY}} \mu(d\xi)\\
&\leq \bar{\knl}\int_\calM\|v(\xi)\|_\YY \mu(d\xi) \leq \bar{\knl}\sqrt{\mu(\calM)}\|v\|_{L^2(\calM,\YY)}.
\end{align*}

Since $\Knl$ is a Mercer kernel,  we know that $\vR_\calM\hookrightarrow C(\calM,\YY)$, which further implies the continuous embedding  
\[
\vR_\calM(\calM,\YY) \overset{\mathcal{I}}{\hookrightarrow} L^2_\mu(\calM,\YY). 
\]
We  can identify the integral operator $L\equiv \mathcal{I}^*$, where $\mathcal{I}$ is the canonical injection of $\vR_\calM\overset{\mathcal{I}}{\hookrightarrow}L^2_\mu(\calM,\YY)$. This follows directly since for every $v\in L^2(\calM,\YY)$ and $h\in \vR_\calM\hookrightarrow L^2_\mu(\calM,\YY)$ we have 
\begin{align*}
\left (Lv,h \right)_{\vR_\calM} & = \left (\int_\calM \Rnl_{\calM}(\cdot,\xi)v(\xi)\mu(d\xi),h(\cdot) \right )_{\vR_\calM} \\
& = \int_\calM \left ( \Rnl_{\calM,\xi} v(\xi),h(\cdot)\right)_{\vR_\calM} \mu(d\xi)\\
&= \int_\calM \left (v(\xi),h(\xi) \right)_\YY \mu(d\xi)=(v,h)_{L^2(\calM,\YY)}=(v,\mathcal{I}h)_{L^2(\calM,\YY)}.
\end{align*}

It is also clear from the simple calculation 
\begin{align*}
    \left (Lv(\xi),y \right)_\YY &= \left ( \int_\calM \Rnl_\calM(\xi,\eta)v(\eta)\mu(d\eta),y\right )_\YY = \left (v , \Rnl_\calM^T(\xi,\cdot)y\right)_{L^2(\calM,\YY)}, 
\end{align*}
 that the feature mappings 
\begin{align*}
    \Psi(\xi)&\triangleq \Rnl_\calM^T(\xi,\cdot) \in \calL(\YY,L^2(\calM,\YY)),\\
    \Psi^*(\xi)&\triangleq (Lv)(\xi)=\int_\calM \Rnl_\calM(\xi,\eta)v(\eta)\mu(d\eta)\in \calL(L^2(\calM,\YY),\YY),
\end{align*}
are bounded linear operators. 

\subsection{General Global Pointwise Error Bounds in Maneuver vRKHSs}
The above observations enable the following error bound.
\begin{theorem}
    \label{th:doublingtrick}
    Let $\Knl:\XX \times \XX \to \calL(\YY)$ be an admissible, operator-valued kernel that defines the vRKHS $\vK$, and suppose $\Knl$ is uniformly bounded on the diagonal by a constant $\bar{\knl}$. Further suppose that   
    $\calM\subset \XX$ is a compact, connected, smooth $\ell$-dimensional Riemannian submanifold that is regularly embedded in $\XX$. Denote by $\vR_\calM(\mathcal{M},\YY)$ the vRKHS that is defined in terms of the restricted kernel $\Rnl_\calM=\Knl|_{\mathcal{M}\times \mathcal{M}}$, and let $\vK_\calM=\Ext_\calM(\vR_\calM)\subset \vK$ be the maneuver vRKHS.  We assume that the restricted kernel $\Rnl_\Omega$ is a Mercer kernel.   
    Any  $f\in \vK_\calM(\XX,\YY)$ has the unique representation $f=\Ext_\calM r$ for some  $r\in \vR_\calM(\calM,\YY)$.  There is a constant $C>0$ such that for any $r\in \Rnl_\calM(\calM,\YY)$ that satisfies the regularity condition $r=Lv$ for some $v\in L^2_\mu(\calM,\YY)$,  we have the \textit{global pointwise error} bound 
    \begin{align*}
        \|E_x(I-\bm{\Pi}_N)f\|_\YY \leq C \sup_{\xi\in \calM}\bar{\Pwr}_{\vR_N}(\xi)\|r\|_{\vR_\calM(\calM,\YY)}\|v\|_{L^2(\calM,\YY)} \quad \quad \text{ for all } x\in \XX, 
    \end{align*}
    where $E_x$ is the evaluation operator on $\vK$ and $\bar{\Pwr}_{\vR_N}$ is the operator-valued power function of the subspace $\vR_N$ in $\vR_\calM$ given in either Equation \ref{eq:PWR_2} or Equation \ref{eq:PWR_infty}. 
\end{theorem}
\begin{proof}
    The norm on $\vK$ dominates the pointwise norm, so we know 
    \begin{align*}
        \|E_x(I-\Pi_N)f\|_\YY & \leq \bar{\knl}\|(I-\Pi_N)\Ext_\calM r\|_\vK \\
        &= \bar{\knl}\|\Ext_\calM(I-\tilde{\Pi}_N)r\|_\vK \\
        &= \bar{\knl}\|(I-\tilde{\Pi}_N)r\|_{\vR_\calM}.
    \end{align*}
    Here we have used the fact that the boundedness of the operator kernel on the diagonal ensures that $\|E_x\|\leq \bar{\Knl}$, the approximations satisfy $\Ext_\calM \tilde{\Pi}_N=\Pi_N \Ext_\calM$ since the centers $\Xi_N\subset \calM\subset \XX$, as well as the fact that the extension operator  $\Ext_\calM\triangleq T_\calM^*:\vR_\calM\to \vK_\calM$ is an isometry. Next, we consider the string of inequalities 
    \begin{align*}
        \|(I-\tilde{\Pi}_N)r\|_{\vR_\calM}&= \left ((I-\tilde{\Pi}_N)r,\ (I-\tilde{\Pi}_N)r \right)_{\vR_\calM},   \\
        &= \left ((I-\tilde{\Pi}_N)r, Lv\right)_{\vR_\calM}, \\
        &=\left ((I-\tilde{\Pi}_N)r, v\right)_{L^2(\calM,\YY)}\\
        &\leq \|(I-\tilde{\Pi}_N)r\|_{L^2(\calM,\YY)}\|v\|_{L^2(\calM,\YY)}.
    \end{align*}
    Finally, we use the analysis of the error in terms of the power function. We have 
    \begin{align*}
        \|(I-\tilde{\Pi}_N)r\|_{L^2(\calM,\YY)}^2&= \int_\calM \|E_\xi(I-\tilde{\Pi}_N)r\|_\YY^2 \mu(d\xi), \\
        &\leq \int_\calM \left (\sup_{\eta\in \calM} \bar{\Pwr}_{\vR_N}(\eta)\|r\|_{\vR_\calM(\calM,\YY)}\right )^2 \mu(d\xi)\\
        &\leq \mu(\calM)\left (\sup_{\eta\in \calM} \bar{\Pwr}_{\vR_N}(\eta)\|\right )^2\|r\|_{\vR_\calM(\calM,\YY)}^2, 
    \end{align*}
    where we have chosen the  power function $\bar{\Pwr}_{\vR_N}\triangleq \bar{\Pwr}_{2,\vR_N}$ in Equation \ref{eq:PWR_2}. 
    The theorem is proven by combining these two above bounds. Choosing instead the power function $\bar{\Pwr}_{\vR_N}=\bar{\Pwr}_{\infty,\vR_N}$ as in Equation \ref{eq:PWR_infty} simply introduces an additional factor of $\sqrt{m}$ in the final coefficient $C>0$ in the theorem. 
\end{proof}
\begin{remark}
    The above bounds give sufficient conditions to ensure  uniform global pointwise error bounds for any $f\in \vK_\calM\subset \vK$. The fidelity of approximation is ensured by checking the power function $\bar{\Pwr}_{\vR_N}(\xi)$ over the restricted manifold, which is used to infer the uniform bounds on the globally defined $f\in \vK_\calM\subseteq \vK$.  
\end{remark}
\begin{remark}
    Note carefully that choosing the centers  $\Xi_N\subset \calM$ increasingly dense in $\calM$ will make the power function $\bar{\Pwr}_{\vR_N}=\bar{\Pwr}_{2,\vR_N}$, given by 
    \begin{align*}
        \bar{\Pwr}_{\vR_N}(\xi)\triangleq \sqrt{\left \|\Rnl_\calM(\xi,\xi)-\Rnl_{\calM,N}(\xi,\xi) \right \|_{2,2} } \quad \quad \text{ for all } \xi \in \calM,
    \end{align*}
    have more zeros on $\calM$, ensuring the decrease of the pointwise error $\|E_\xi(I-\tilde{\Pi}_N)r\|_\YY$ for $\xi\in \calM$. However,  having dense  samples $\Xi_N$ in $\mathcal{M}$ will not imply  that the globally defined  power function $\bar{\Pwr}_{\vK_N}$ of the subspace $\vK_N$ in $\vK$ is small, where 
    \[
    \bar{\Pwr}_{\vK_N}(x)= \bar{\Pwr}_{\vK_N,\vK}(x)= \sqrt{\left \| \Knl(x,x)-\Knl_{N}(x,x) \right \|}, \quad \quad \text{ for all } x\in \XX. 
    \]
    Note that $\bar{\Pwr}_{\vR_N}=\bar{\Pwr}_{\vK_N}|_{\calM}$, and  since $\Rnl=\Knl|_{\calM\times \calM}$, so we do expect that $\bar{\Pwr}_{\vK_N}(x)=\bar{\Pwr}_{\vK_N,\vK}(x)$ to be small for $x\in \calM\subset \XX$.  
\end{remark}

\section{Fill Distances and Maneuver vRKHS: A First Result}
\label{sec:fill_distance_maneuver}

The bounds above are useful and general in that they are expressed in terms of the operator-valued power function $\Pwr_{\vR_N}(\xi)$, 
 which is defined for any vRKHS $\vR_\Omega$. It  can be computed for any $\xi\in \calM$ once a (candidate) set of centers $\Xi_N\subset \calM$ is selected. These bounds can then be used for either \textit{a priori} or \textit{a posteriori} error estimates. They fall  short, however, of providing enough information to make complexity estimates as described in Equation \ref{eq:form_complexity} in the introduction.

In this section we specialize the above bounds to a setting that enables such scaling arguments. The analysis in this section, in particular, enables such bounds for operator-valued kernels $\Knl$ that are diagonal and have the form 
\begin{align}
    \Knl(x_1,x_2)=\knl(x_1,x_2) I_m \in \calL(\YY) \quad \quad \text{ for all } x_1,x_2 \in \XX\triangleq \RR^n 
\end{align}
for some scalar-valued kernel $\knl:\XX \times \XX \to \RR$ where $I_m$ is the identity matrix in $\RR^{m\times m}$. The kernel $\knl$ defines a scalar-valued RKHS $\sH(\XX,\RR)$, and hence $\vH(\XX,\YY)\triangleq (\sH(\XX,\RR))^m$. 

\begin{theorem}
\label{th:diagonal_C2s}
    Let $\knl:\XX\times \XX \to \RR$ be a positive definite kernel that is uniformly bounded on the diagonal by a constant $\bar{\knl}>0$,  $\calM\subset \XX$ be a smooth, connected, compact, $\ell$-dimensional Riemannian manifold that is regularly embedded in $\XX$, and suppose that the restricted kernel $\rnl_\calM\triangleq \knl|_{\calM\times \calM}\in C^{2\bar{s}}(\calM \times \calM, \RR)$ for an integer smoothness $\bar{s}\geq 1$.  Denote by $\vK$ the vRKHS induced by the diagonal operator-valued kernel $\Knl\triangleq \knl I_m$ where $I_m$ is the identity matrix on $\RR^m$, define the restricted vRKHS $\vR_\calM\triangleq T_\calM(\vK)$, and the maneuver space $\vK_\calM\triangleq \Ext_\calM(\vR_\calM)$. For any $f\in \vK_\calM$ there is a unique $r\in \vR_\calM$ such that $f=\Ext_\calM r$. There is a constant $C>0$ such that we have the global bound 
    \[
    \|E_x(I-\bm{\Pi}_N)f\|_\YY\leq C h^{\bar{s}}_{\Xi_N,\calM} \|r\|_{\vR_\calM} \|v\|_{L^2_\mu(\calM,\YY)}
    \]
    for all $f=\Ext_\calM r\in \vK_\calM$ that satisfy the regularity condition $r=Lv$ for some $v\in L^2_\mu(\calM,\YY)$. 
\end{theorem}
\begin{proof}
    We only prove the case above when $m=1$, since the general case follows from simple component-wise considerations for the diagonal operator-valued kernels. When $m=1$, Theorem 17.21 of \cite{wendland} ensures that 
    \[
    |E_\xi(I-\tilde{\Pi}_N)r|\leq Ch^{\bar{s}}_{\Xi_N,\calM} \|r\|_{\vR_\calM} \quad \quad \text{ for all } \xi\in \calM,  
    \]
    which can be integrated over the manifold to obtain 
    \[
    \|(I-\tilde{\Pi}_N)r\|_{L^2_\mu(\calM,\YY)}\leq C\sqrt{\mu(\calM)}h^{\bar{s}}_{\Xi_N,\calM}\|r\|_{\vR_\calM}. 
    \]
    Arguing as in the proof of Theorem \ref{th:doublingtrick}, we also have 
    \begin{align*}
    \|(I-\tilde{\Pi}_N)r\|_{\vR_\calM} &\leq \|(I-\tilde{\Pi})f\|_{L^2_\mu(\calM,\YY)} \|v\|_{L^2_\mu(\calM,\YY)}\\
    &\leq C\sqrt{\mu(\calM)} h^{\bar{s}}_{\xi_N,\calM} \|r\|_{\vR_\calM}\|v\|_{L^2_\mu(\calM,\YY)}. 
    \end{align*}
    Finally, using the uniform bound $\|E_x\|\leq \bar{\knl}$ and the fact that $\Ext_\calM:\vR_\calM\to \vK_\calM$ is an onto isometry, we conclude
    \begin{align*}
    \|E_x(I-\bm{\Pi_N})f\|_\YY&\leq \bar{\knl}\|\Ext_\calM(I-\tilde{\Pi}_N)r\|_{\vR_\calM} = \bar{\knl}\|(I-\tilde{\Pi}_N)r\|_{\calR_{\calM}} \quad \quad \text{ for all } x\in \XX. 
    \end{align*}
    Combining this inequality with the last one above completes the proof of the theorem. 
\end{proof}

This theorem is attractive in its simplicity.  However, it leaves out an important detail that is crucial for applications. The theorem above relies on finding a global scalar-valued kernel $\knl:\XX\times \XX \to \RR$, which when restricted to the manifold $\calM$, generates a reproducing kernel on the manifold that is of class $C^{2\bar{s}}(\calM\times\calM, \RR)$. No advice on how precisely this can be accomplished is described above, nor is it given in the discussion of Theorem 17.21 in \cite{wendland}.

This is the topic we consider next, how to achieve such bounds in practical cases. 

\subsection{Interpolation and Projection Bounds in Sobolev Spaces}
\label{sec:sobolev_bounds}

The construction in this section  is carried out for a certain popular, well-known class of radial kernels $\knl(x,y)\triangleq \eta(x-y)$ with the radial function $\eta:\RR^n \to \RR^+$. We say that the Fourier transform $\hat{\eta}(\omega)$ of the radial function $\eta$ has algebraic decay of order $s>n/2$ if there are two constants $C_1,C_2>0$ such that 
\begin{align}
C_1 (1+\|\omega\|^2_2)^{-s} \leq \|\hat{\eta}(\omega)\|^2 \leq C_2 (1+\|\omega\|^2_2)^{-s} \quad \text{ for all } \omega\in \RR. \label{eq:fourier_decay_r}
\end{align}
In this case it is known, see Corollary 10.13 of \cite{wendland},  that the scalar-valued RKHS $\sH(\RR^n,\RR)$ defined by the kernel $\knl$ is equivalent to the Sobolev space 
\begin{align*}
    \sH(\RR^n,\RR) \approx \sW^{s,2}(\RR^n,\RR).
\end{align*} 
There are a significant and large collection of standard radial scalar-valued kernels that have a Fourier transform with algebraic decay. These include the Sobolev-Matern kernel $\knl_s$ of smoothness $s>0$, the Wendland kernels $\knl_{\sigma(n,s)}$, and the Abel kernels for $s=(n+1)/2$. See Table 11.1 in \cite{wendland}, Table 1 in \cite{schaback94}, or Equation 14 in \cite{de2014learning}. 

If $\Omega\subset \RR^n$ is a connected open set with a Lipschitz boundary, it is well-known that taking the restrictions of functions in $\sH(\RR^n,\RR)$ to obtain the native space $\sH(\Omega,\RR)$ yields the corresponding Sobolev space 
\[
\sH(\Omega,\RR)\approx \sW^s(\Omega,\RR).
\]
See Corollary 10.48 in \cite{wendland}. It is important to note here the smoothness index $s$ for the Sobolev space of functions over $\Omega$ is the same as the smoothness index $s$ for the Sobolev space of functions defined on all of $\RR^n$. However, we are interested in this paper in the case when the subset over which restrictions are taken is a compact, connected, smooth $\ell$-dimensional Riemannian submanifold of $\RR^n$. Such a manifold has an empty interior in $\RR^n$, so it has Lebesgue measure zero in $\RR^n$. The simple situation whereby the restriction operation for a nice domain $\Omega$ preserves the Sobolev smoothness $s$ is no longer generally true for restrictions to such a manifold $\calM$. 

Instead, in this paper we use  results from  \cite{fuselier2012} that establishes that restrictions to $\calM$ of the radial kernels above  generate Sobolev spaces $\sW^{\bar{s}}(\calM,\RR)$ having \textit{reduced smoothness $\bar{s}<s$}, with $\bar{s}\triangleq s-(n-\ell)/2$.  That is, if the kernel $\knl:\RR^n \times \RR^n\to \RR$ is defined in terms of the radial function $\eta$ with a Fourier transform $\hat{\eta}(\omega)$ having algebraic decay $s>n/2$, then 
\begin{align*}
    \sR_{\calM}(\calM,\RR)\approx \sW^{s-(n-\ell)/2}(\calM,\RR) \triangleq \sW^{\bar{s}}(\calM,\RR). 
\end{align*}
The utility of this observation is the following Theorem 11 of \cite{fuselier2012}.
\begin{theorem}[The Many-Zeros Theorem, simplified from \cite{fuselier2012}]
\label{th:many_zeros}
Let $\calM\subset \RR^n$ be a $\ell$-dimensional, compact, connected, smooth Riemannian manifold that is regularly embedded in $\RR^n$, suppose the the radial kernel $\knl:\RR^n \times \RR^n \to \RR$ is defined in terms of the radial function $\eta$ that satisfies the algebraic decay condition in Equation \ref{eq:fourier_decay_r} for $s>n/2$.  Define  the reduced smoothness $\bar{s}\triangleq s-(n-\ell)/2$. Then there is a constant $C>0$ such that for all sufficiently fine set of centers $\Xi_N\subset \calM$ and $t \in [0,\lceil \bar{s}\rceil]$, we have 
\begin{align*}
    \|(I-\tilde{\Pi}_N)f\|_{\sW^{t}(\calM,\RR)} \leq C h^{\bar{s}-t}_{\Xi_N,\calM}\|f\|_{\sR_\calM(\calM,\RR)}
\end{align*}
for any $f\in \sR_{\calM}(\calM,\RR)$. 
\end{theorem}

\begin{remark}
The phrase ``sufficiently fine'' refers to the fact that the inequality above holds for all fill distances less than some critical value $h_{0}>0$. Expressions for the critical value of the fill distance can be found in \cite{fuselier2012,narcowich2010,narcowich2012}, but we do not use them in this paper. In the discussions below we let an integer $N_0>0$ denote a number of centers sufficient to ensure that $\Xi_{N_0}$ is sufficiently fine in the set of interest. 
\end{remark}

\begin{remark}The above version of the Many Zeros Theorem for scalar-valued RKHS $\sR_\calM(\calM,\RR)$ has a simple extension to the vRKHS $\vK=\sK^m$ that is defined in terms of the diagonal operator-valued kernel $\Knl\triangleq \knl I_m$. We then have
\begin{align*}
    \|(I-\tilde{\bm{\Pi}}_N)f\|_{\vW^{t}(\calM,\RR^m)} \leq C h^{\bar{r}-t}_{\Xi_N,\calM}\|f\|_{\vR_\calM(\calM,\RR^m)} 
\end{align*}
for any $f\in \vR(\calM,\RR^m)$. 
\end{remark}

\subsection{Refined Pointwise Global Error Bounds}
This section refines the general error bounds summarized in Theorem \ref{th:doublingtrick} by exploiting the many-zeros theorem for scalar-valued RKHS. 
\begin{theorem}
    \label{th:pointwiserefined}
    Let the operator-valued kernel $\Knl:\RR^n\times \RR^n\to \calL(\RR^m)\triangleq \knl I_m$ generate the global vRKHS $\vK(\XX,\YY)$ for a scalar-valued kernel $\knl$, denote by $\vR_{\calM}(\calM,\YY)$ the restricted vRKHS $\vR(\calM,\YY)=T_\calM(\vK(\XX,\YY))$, define the global maneuver vRKHS as $\vK_\calM \triangleq \Ext_\calM(\vR_\calM)$, let the scalar-valued kernel satisfy the hypotheses of Theorem \ref{th:many_zeros}, and assume the hypotheses of Theorem \ref{th:doublingtrick} hold.    Then there is a constant $C>0$ such that for all sufficiently fine sets of centers $\Xi_N\subset \calM$  we have  the \textit{global pointwise error bound} for any function in the maneuver space $f\in \vK_\calM$ given by 
    \begin{align*}
        \|E_x(I-\bm{\Pi}_N)f\|_\YY \leq Ch^{\bar{s}}_{\Xi_N,\calM}\|r\|_{\vR_\calM(\calM,\RR^m)}\|v\|_{L^2(\calM,\RR^m)} \quad \quad \text{ for all } x\in \XX,
    \end{align*}
 where $f=\Ext_\calM r$ for the unique  $r\in \vR_\calM(\calM,\RR^m)$. 
\end{theorem}
\begin{proof}
    From the proof of Theorem \ref{th:doublingtrick}, we know that 
    \begin{align*}
        \|E_x(I-\bm{\Pi}_N)f\|_\YY &\leq \|(I-\tilde{\bm{\Pi}})r\|_{\vR_{\calM}(\calM,\YY)} \\
        &\leq \|(I-\tilde{\bm{\Pi}}_N)r\|_{L^2_\mu(\calM,\YY)}\|v\|_{L^2_\mu(\calM,\YY)}.
    \end{align*}
    But we can now apply the Many Zeros Theorem for the choice of $t=0$ to obtain 
    \begin{align*}
        \|E_x(I-\bm{\Pi}_N)f\|_\YY &\leq C h^{\bar{s}}_{\Xi_N,\calM}\|r\|_{\vR_\calM(\calM,\RR^m)}\|v\|_{L^2_\mu(\calM,\RR^m)},
    \end{align*}
    which completes the proof. 
\end{proof}

The above bound can now be used to define the desired computational complexity estimates \textit{for the approximation of functions}, which we use subsequently in Section \ref{sec:nonparametric_and_jackson_inequalities} to derive corresponding performance bounds for controllers. It is known, see for example \cite{wenzel2023analysis},  that for quasiuniform samples $\Xi_N\subset \Omega$ in a parallelopiped, which we can take as  $\Omega\triangleq [0,1]^\ell\subset \RR^\ell$ without loss of generality,  we have that the number of centers $N$ scales like 
\[
N=N(\Xi_N,\Omega) \sim \frac{1}{h_{\Xi_N,\Omega}^{\ell}}=h_{\Xi,\Omega}^{-\ell}.
\]
By passing from $\calM$ to $\RR^\ell$ via coordinate charts, the same scaling is true for quasiuniform centers in a compact manifold $\calM$, that is, 
\[
N=N(\Xi_N,\calM) \sim \frac{1}{h_{\Xi_N,\calM}^{\ell}}=h_{\Xi_N,\calM}^{-\ell}.
\]
See for example the proof of Theorem 5.6 of \cite{narcowich2012}, or the discussion in Section \ref{sec:point_sets_distances}.  These show that the number of centers  in $\Xi_N\subset \calM$ scales like the $r^{-\ell}_{\Xi_N}$ where $r_{\Xi_N}$ is the minimal separation radius of the set of centers $\Xi_N$.

When we choose the performance target for the pointwise approximation of a function in $\vK_\calM$ to be 
\[
\|E_x(I-\Pi_N)f\|_\YY \sim  \epsilon \sim h^{\bar{s}}_{\Xi_N,\calM}, 
\]
it follows from the bound that the number of quasiuniform  centers in $\calM$ scales like 
\[
N=N(\calM)\sim \frac{1}{\epsilon^{\ell/\bar{s}}}. 
\]

\section{Nonparametric Adaptive Control and Maneuver vRKHSs}
\label{sec:nonparametric_and_jackson_inequalities}

In this section we study a popular generalization of the model problem introduced in Equation \ref{eq:model1}. Overall, the analysis of the model problem in this section is largely based on the proof of Theorem 6.4 in \cite{boffi2022nonparametric}. A few changes are made to structure the problem in the standard language of MRAC and to  emphasize  how methods  based on smoothed deadzones can yield nonparametric adaptive controllers that are AAO.    

  We formulate the problem  by further breaking down the functional uncertainty  into parametric parts and nonparametric parts in the equation 
\begin{align}
    \dot{x}(t)=Ax(t)+B\Lambda(\mu(t)+\Theta^T\Phi(x(t))+E_{x(t)}f). \label{eq:model2}
\end{align}
Again,  in this equation the state $x(t)\in \XX\triangleq \RR^n$, the control $\mu(t)\in \RR^m\triangleq \UU$,  and the nonparametric uncertainty $f:\XX \to \UU$.  But now  we have also introduced a  parametric functional uncertainty  $(\Theta^T\Phi)(\cdot):\XX \to \UU$ that is expressed as the product of a (true) parameter matrix $\Theta\in \RR^{p\times m}$ and a  known regressor vector $\Phi(x)\in \mathbb{P}\triangleq \RR^p$. 

The goal is to derive a control input $\mu(t,x)\in \UU$ such that we ultimately track the following reference system
\[
    \dot{x}_{ref}(t)=A_{ref} x_{ref}(t)+B_{ref} r(t),
\]
where, $A_{ref}$ is Hurwitz. Similarly to classical parametric MRAC theory, the adaptive controller is interpreted as an approximation of the  \textit{nonparametric} feedback controller   
\begin{align}
    \mu^\ast(t)=K_x^T x(t)+ K_r^T r(t)-\Theta^T\Phi(x(t))-E_{x(t)}f, \label{eq:ideal_nonparametric_controller}
\end{align}
where the matrices $K_x\in \RR^{n\times m}$, $K_r\in \RR^{m\times m}$ are ideal linear gain matrices that satisfy the following matching conditions
\begin{align*}
    A_{ref}=A+B\Lambda K_x^T,\quad B_{ref}=B\Lambda K_r
\end{align*}
and $\Lambda\in \RR^{m\times m}$. 

Even if the true real parameters $K_x,K_r,\Theta$ were known,  the nonparametric feedback controller above is  still  not realizable owing to the nonparametric uncertainty $f$. Since we only know that $f\in \vK$, this in general function requires an infinite number of coefficients to represent. As suggested in the previous discussions, we represent approximations of the uncertainty $f$ that are based on finite dimensional space 
\begin{equation}
\vK_N\triangleq \text{span} \{\vK_{\xi_i}y \ | \ \xi_i \in \Xi_N, y\in \YY\}\subset \vK \label{eq:definition_KN}
\end{equation}
for centers $\xi_i\in \Xi_N\subset \calM\subset \XX$. 
Approximations $\bm{\Pi}_Nf$ are constructed using the $\vK$-orthogonal projection of $\vK$ onto $\vK_N$. 

The specific algorithms derived in this section are all examples of smoothed deadzone methods as defined in \cite{boffi2022nonparametric}, Definition 6.1, which we recall verbatim here.
\begin{defn}
    Let $\Delta>0$. A continuously differentiable function $\sigma:\RR^+\to \RR$ is called a $\Delta$-admissible (smoothed) deadzone function if 
    \begin{enumerate}
        \item $0\leq \sigma$ and $\sigma(x)=0$ for all $x\in [0,\Delta]$.
        \item $0\leq \sigma'$ and $\sigma'(x)=0$ for all $x\in [0,\Delta]$.
        \item $\sigma'$ is locally Lipschitz continuous.  
    \end{enumerate}
\end{defn}

\noindent With such a smoothed deadzone function $\sigma$, we define the following learning laws:
\begin{align}
    \begin{cases}
        \dot{\tilde{K}}_x=-{\sigma}'({e}^T{Pe})\cdot \Gamma_x x  e^T P B,\\
        \dot{\tilde{K}}_r=-{\sigma}'(e^TPe)\cdot \Gamma_r r e^T P B, \\
        \dot{\tilde{\Theta}}={\sigma}'({e}{Pe})\cdot \Gamma_\Theta \Phi( x) e^T P B,\\
        \dot{\hat{f}}_N(t,\cdot)={\sigma}'({e}{Pe})\cdot \Gamma_f \bm{\Pi}_N \Knl_{x(t)}(\cdot)B^T P e(t), 
    \end{cases} \label{eq:final_learning_laws}
\end{align}
and the realizable or approximate feedback controller is defined to be 
\begin{align}
    \mu(t)=\hat{K}_x^T(t) x(t)+ \hat{K}_r^T(t) r(t)-\hat{\Theta}^T(t)\Phi(x(t))-E_{x(t)}\hat{f}_N(t,\cdot). \label{eq:realizable_nonparametric_controller}
\end{align}
The first three learning laws in Equation \ref{eq:final_learning_laws} are standard  update laws from  classical parametric MRAC theory. The last equation in terms of $\bm{\Pi}_N\Knl_{x(t)}$ for the update of the online estimate $\hat{f}_N$ of the functional uncertainty is the standard one used in deterministic reproducing kernel embedding techniques, see \cite{kurdilaBook}. 

In this case  the following theorem is only a slight modification of Theorem 6.4 in \cite{boffi2022nonparametric} to cast the problem in the standard language of MRAC theory and emphasize that the (smoothed) deadzone control method is AAO.  
\begin{theorem} 
\label{th:smoothed_deadzone_AAO}
    For the original system defined in Equation \ref{eq:model2} assume  that the finite dimensional spaces $\vK_N$ are defined   as in Equation \ref{eq:definition_KN} for a set of centers $\Xi_N\subset \XX$, the positive constant $R>0$ is defined in Equation \ref{eq:defn_R}, the learning laws are given by Equations \ref{eq:final_learning_laws}, and the feedback controller is selected as in Equation \ref{eq:realizable_nonparametric_controller}, and the deadzone size $\Delta$ satisfies 
    \begin{align*}
    \Delta  > \bar{\Delta}_N\triangleq  R\|PB\|\lambda_{min}(Q)\lambda_{min}(P)\sup_{\|\xi\|_\XX\leq \bar{R}}\|E_{\xi}(I-\bm{\Pi}_N)f\|_\UU
    \end{align*}
    for $\bar{R}>0$ defined in Equation \ref{eq:defn_barR}. Then for any system in Equation \ref{eq:model2} with  $f\in \vK$,  we have 
    \[
    \limsup_{t\to \infty} \|x(t)-x_r(t)\|_\XX \leq \Delta. 
    \]
    The  nonparametric control method defined by the family of deadzones  $\{\Delta_N\}_{N\geq N_0}$ is AAO, when   $\Delta_N\triangleq (1+\delta)\bar{\Delta}_N$, $\delta>0$ is a small constant,  and   $\bar{\Delta}_N$ is defined above.   
\end{theorem}

\begin{proof}
The outline of the  proof below is included for completeness and to emphasize that the nonparametric adaptive control method defined in terms of $\bm{\Pi}_N$ is AAO. See  Theorem 6.4 of \cite{boffi2022nonparametric} for the details in  a slightly different setting. 

We consider the Lyapunov function candidate
\begin{align*}
V(e,\tilde{K}_x,\tilde{K}_r,\tilde{\Theta},\tilde{f}) &= \sigma ({e}{Pe}) + \iptr{\tilde{K}_x}{\Gamma_x^{-1}\tilde{K}_x\Lambda}+\iptr{\tilde{K}_r}{\Gamma_r^{-1}\tilde{K}_r\Lambda}\\ &\hspace*{.2in}+\iptr{\tilde{\Theta}}{\Gamma_\Theta^{-1}\tilde{\Theta}\Lambda}+\ipvK{\tilde{f}}{\Gamma_f^{-1}\Lambda \tilde{f}}. 
\end{align*}
When we take the derivative of this function along the trajectories of the closed loop equations, we generate the following sequence of inequalities:
\begin{align*}
    \dot{V}&= \sigma' (\ipRd{e}{Pe}) (\ipRd{\dot{e}}{Pe}+ \ipRd{e}{P\dot{e}})\\
    &\quad+ \iptr{\dot{\tilde{K}}_x}{\Gamma_x^{-1}\tilde{K}_x\Lambda}+\iptr{\tilde{K}_x}{\Gamma_x^{-1}\dot{\tilde{K}}_x\Lambda}+\iptr{\dot{\tilde{K}}_r}{\Gamma_r^{-1}\tilde{K}_r\Lambda}+\iptr{\tilde{K}_r}{\Gamma_r^{-1}\dot{\tilde{K}}_r\Lambda}\\
    &\quad+\iptr{\dot{\tilde{\Theta}}}{\Gamma_\Theta^{-1}\tilde{\Theta}\Lambda}+\iptr{\tilde{\Theta}}{\Gamma_\Theta^{-1}\dot{\tilde{\Theta}}\Lambda}+\ipvK{\dot{\tilde{f}}}{\Gamma_f^{-1}\Lambda\tilde{f}}+\ipvK{\tilde{f}}{\Gamma_f^{-1}\Lambda \dot{\tilde{f}}},\\
    \\
    &= \sigma' (\ipRd{e}{Pe})\left(\ipRd{[A_{ref} e + B\Lambda(-\tilde{K}_x^T x - \tilde{K}_r^T r + \tilde{\Theta}^T(t)\Phi( x(t))+ E_{x}\tilde{f})]}{Pe}+\right.\\
    &\quad \left.+ \ipRd{e}{P[A_{ref} e + B\Lambda(-\tilde{K}_x^T x - \tilde{K}_r^T r +\tilde{\Theta}^T(t)\Phi( x(t))+ E_{x}\tilde{f})]}\right) + \\
    &\quad - 2 \iptr{\dot{\hat{K}}_x}{\Gamma_x^{-1}\tilde{K}_x\Lambda} -2\iptr{\dot{\hat{K}}_r}{\Gamma_r^{-1}\tilde{K}_r\Lambda}-2\iptr{\dot{\hat{\Theta}}}{\Gamma_\Theta^{-1}\tilde{\Theta}\Lambda}-2\ipvK{\dot{\hat{f}}}{\Gamma_f^{-1} \Lambda  \tilde{f}},\\
    \\
\end{align*}
\begin{align*}
    &= \sigma' (\ipRd{e}{Pe}) \left(\ipRd{A_{ref} e}{Pe}+\ipRd{e}{P A_{ref} e}\right.\\
    &\quad -\ipRd{ B\Lambda\tilde{K}_x^T x }{Pe}-\ipRd{e}{PB\Lambda\tilde{K}_x^T x} -\ipRd{ B\Lambda\tilde{K}_r^T r }{Pe}-\ipRd{e}{PB\Lambda\tilde{K}_r^T r}\\
    &\quad +\ipRd{ B\Lambda\tilde{\Theta}^T \Phi( x) }{Pe}+\ipRd{e}{PB\Lambda\tilde{\Theta}^T \Phi( x)} + \ipRd{ B\Lambda E_{x}\tilde{f} }{Pe}+\ipRd{e}{PB\Lambda E_{x}\tilde{f}})\\
    &\quad  - 2 \iptr{\dot{\hat{K}}_x}{\Gamma_x^{-1}\tilde{K}_x\Lambda} -2\iptr{\dot{\hat{K}}_r}{\Gamma_r^{-1}\tilde{K}_r\Lambda}-2\iptr{\dot{\hat{\Theta}}}{\Gamma_\Theta^{-1}\tilde{\Theta}\Lambda}-2\ipvK{\dot{\hat{f}}}{\Gamma_f^{-1} \Lambda \tilde{f}},\\
    \\
    &= \sigma' (\ipRd{e}{Pe})\left(\ipRd{e}{A_{ref}^T P e}+\ipRd{e}{P A_{ref} e}-2\ipRd{\Lambda \tilde{K}_x^T  }{ B^T Pe x^T}-\right.\\
    &\quad\left.- 2 \ipRd{\Lambda \tilde{K}_r^T }{ B^T Pe r^T} +2 \ipRd{\Lambda \tilde{\Theta}^T }{ B^T Pe \Phi( x)^T}+ 2\ipRd{\Lambda E_{x}\tilde{f} }{ B^T Pe}\right)-\\
    &\quad - 2 \iptr{\dot{\hat{K}}_x}{\Gamma_x^{-1}\tilde{K}_x\Lambda} -2\iptr{\dot{\hat{K}}_r}{\Gamma_r^{-1}\tilde{K}_r\Lambda}-2\iptr{\dot{\hat{\Theta}}}{\Gamma_\Theta^{-1}\tilde{\Theta}\Lambda}-2\ipvK{\dot{\hat{f}}}{\Gamma_f^{-1} \Lambda \tilde{f}},\\
    \\
    &= -\sigma' (\ipRd{e}{Pe})\ipRd{e}{Qe}-\\
    &\quad-2\left(\sigma' (\ipRd{e}{Pe})\ipRd{ \tilde{K}_x \Lambda }{x  e^T P B }+\iptr{\tilde{K}_x\Lambda}{\Gamma_x^{-1}\dot{\hat{K}}_x}\right)-\\
    &\quad-2\left(\sigma' (\ipRd{e}{Pe})\ipRd{ \tilde{K}_r \Lambda }{r  e^T P B }+\iptr{\tilde{K}_r\Lambda}{\Gamma_r^{-1}\dot{\hat{K}}_r}\right)+\\
    &\quad + 2\left(-\sigma' (\ipRd{e}{Pe})\ipRd{ \tilde{\Theta} \Lambda }{\Phi( x)  e^T P B }+\iptr{\tilde{\Theta}\Lambda}{\Gamma_\Theta^{-1}\dot{\hat{\Theta}}}\right)+\\
    &\quad+2\left(-\sigma' (\ipRd{e}{Pe})\ipRd{ \Lambda \tilde{f} }{E^*_{ x}  B^T Pe}+\ipvK{\Lambda \tilde{f}}{\Gamma_f^{-1}\dot{\hat{f}}}\right),\\
    \\
    &= -\sigma' (\ipRd{e}{Pe})\ipRd{e}{Qe}-\\
    &\quad-2\left(\ipRd{ \tilde{K}_x \Lambda }{\sigma' (\ipRd{e}{Pe}) x  e^T P B }+\iptr{\tilde{K}_x\Lambda}{-\Gamma_x^{-1}\sigma' (\ipRd{e}{Pe})\cdot \Gamma_x x  e^T P B}\right)-\\
    &\quad-2\left(\ipRd{ \tilde{K}_r \Lambda }{\sigma' (\ipRd{e}{Pe}) r  e^T P B }+\iptr{\tilde{K}_r\Lambda}{-\Gamma_r^{-1}\sigma' (\ipRd{e}{Pe})\cdot \Gamma_r r e^T P B}\right)+\\
    &\quad + 2\left(-\ipRd{ \tilde{\Theta} \Lambda }{\sigma' (\ipRd{e}{Pe})\Phi( x)  e^T P B }+\iptr{\tilde{\Theta}\Lambda}{\Gamma_\Theta^{-1}\sigma' (\ipRd{e}{Pe})\cdot \Gamma_\Theta \Phi( x) e^T P B}\right)+\\
    &\quad+2\left(-\ipRd{\Lambda \tilde{f}}{\sigma' (\ipRd{e}{Pe}) E^*_{ x}  B^T Pe}+\ipvK{\Lambda \tilde{f}}{\Gamma_f^{-1}\sigma' (\ipRd{e}{Pe})\cdot \Gamma_f \bm{\Pi}_N \Knl_{ x(t)}(\cdot)B^T P e(t)}\right),\\
    \\
    &=-\sigma' (\ipRd{e}{Pe})\ipRd{e}{Qe}+2\left(\ipRd{ \Lambda \tilde{f}}{\sigma' (\ipRd{e}{Pe}) (-I+\bm{\Pi}_N) E^*_{ x}  B^T Pe}\right)\\
    &=-\sigma' (\ipRd{e}{Pe})\left(\ipRd{e}{Qe}-2\ipRd{\Lambda \tilde{f}}{ (-I+\bm{\Pi}_N) \Knl_{ x(t)}(\cdot)  B^T Pe}\right).
\end{align*}
We next  define  the following uncertainty classes 
\begin{align*}
    \mathcal{C}_x&\triangleq \left \{K_x \ | \ \iptr{{K}_x}{\Gamma_x^{-1}K_x\Lambda} < C_x\right \}, \\
    \mathcal{C}_r&\triangleq \left \{K_r\ | \ \iptr{{K}_r}{\Gamma_r^{-1}{K}_r\Lambda} <C_r\right \},\\
    \mathcal{C}_\Theta&\triangleq \left \{\Theta \ | \ \iptr{{\Theta}}{\Gamma_\Theta^{-1}{\Theta}\Lambda}<C_\Theta \right \},\\
    \calC_f &\triangleq \left \{f \ | \   \ipvK{{f}}{\Gamma_f^{-1}{f}\Lambda}<C_f \right \},
\end{align*}
for the fixed design constants $C_x,C_r,C_\Theta,C_f>0$. Choose some constant $R>0$ that satisfies 
\begin{align}
    \ipRd{e(0)}{Pe(0)} + C_x + C_r + C_\Phi + C_f <\lambda_{min}(P)R^2. \label{eq:defn_R}
\end{align}
We begin by noting that the  governing ODEs have the form 
\[
\dot{Z}(t)=F(t,Z(t))
\]
where $Z(t)\triangleq (e(t),\hat{K}_x(t),\hat{K}_r(t),\hat{\Theta}(t),\hat{f}_N(t,\cdot))\in \ZZ\triangleq\XX \times \RR^{n\times m} \times \RR^{m\times m} \times \RR^{p\times m}\times  \vK_N$, and the right hand side function $F$ is continuous in $(t,Z)$ and locally Lipschitz continuous in $Z$. This implies that there is a maximum interval of existence $[0,T_{max})$ for solutions  of these ODEs,  where $T_{max}$ can be finite or equal to $\infty$. It is also known that if $T_{max}$ is finite, we must have $\|Z(t)\|_{\ZZ}\to \infty$ as $t\to T_{max}$. See  \cite{khalil,khalil2002nonlinear} for a discussion of the existence of solutions of ODEs for locally Lipschitz continuous right hand sides as above. 

Define a time $T_0$ such that
\begin{align}
    T_0=\sup\{T\in[0,T_{\max})\ |\ \|{e(t)}\|_\XX \leq R,\forall t\in[0,T]\} \label{eq:sup_defn_T0}.
\end{align}
The constant $T_0$ is well-defined since 
\[
\|e(0)\|_\XX \leq \frac{1}{\sqrt{\lambda_{min}(P)}}\sqrt{ \ipRd{e(0)}{Pe(0)}}< R.
\]

Suppose that $T_0<T_{\max}$.  Calculation of the derivative of the Lyapunov function along the trajectories for any time $t\in [0,T_0]$ yields 
\begin{align}
    \dot{V}&=\sigma'(e^TPe)\left \{-e^TQe +\|PB\|\|E_{x(t)}(I-\bm{\Pi}_N)f\|_\UU \|e\|_\XX\right \}\label{eq:Vdot_equals}\\
    &\leq \sigma'(e^TPe)\left \{-e^TQe +R\|PB\|\|E_{x(t)}(I-\bm{\Pi}_N)f\|_\UU \right \}\notag \\
    &\leq \sigma'(e^TPe)\left \{-\frac{1}{\lambda_{min}(Q)}\|e\|_\XX^2  +R\|PB\|\|E_{x(t)}(I-\bm{\Pi_N})f\|_\UU \right \}\notag
\end{align}
But we know that over $[0,T_0]$ we also have 
\[
\|x(t)\|_\XX\leq \|x_r(t)\|_\XX + \|e(t)\|_{\XX} \leq \bar{x}_r+R\triangleq \bar{R}.
\]
The above bound can now be written as 
\begin{align}
   \dot{V}\leq \sigma'(e^TPe)\left \{-\frac{1}{\lambda_{min}(Q)}\|e\|_\XX^2  +R\|PB\|\sup_{\|\xi\|_\XX\leq \bar{R}}\|E_{\xi}(I-\bm{\Pi}_N)f\|_\UU \right \} \label{eq:defn_barR}
\end{align}
for any $t\in [0,T_0]$. 
We only need to worry about the right hand side when the tracking error trajectory $t\mapsto e(t)$ is outside the deadzone $\Delta$, which occurs when $e^TPe>\Delta$, since otherwise $\sigma'$ is zero.  This implies that outside the deadzone we have 
\[
-\Delta > -e^TPe \geq -\lambda_{max}(P)\|e\|^2, 
\]
and the derivative of the Lyapunov function becomes 
\begin{align*}
   \dot{V}&\leq \sigma'(e^TPe)\left \{-\frac{1}{\lambda_{min}(Q)\lambda_{min}(P)}\Delta  +R\|PB\|\sup_{\|\xi\|_\XX\leq \bar{R}}\|E_{\xi}(I-\bm{\Pi}_N)f\|_\UU \right \},\\
   &\leq -\sigma'(e^TPe)
   \left \{\frac{\Delta  -R\|PB\|\lambda_{min}(Q)\lambda_{min}(P)\sup_{\|\xi\|_\XX\leq \bar{R}}\|E_{\xi}(I-\bm{\Pi}_N)f\|_\UU}{\lambda_{min}(Q)\lambda_{min}(P)} \right \}, \\
   &\triangleq -\alpha \sigma'(e^TPe).
\end{align*}
For the  properly designed smoothed deadzone we have $V(t)\leq V(0)$ for all $t\in [0,T_0]$. 

But in fact we can argue, similarly to  the proof of Theorem 6.4 of \cite{boffi2022nonparametric}, 
that we must have $T_0=T_{max}$. If not, then the continuity of the solution $t\mapsto Z(t)$ would imply the continuity of the tracking error $t\mapsto e(t)$ over $[0,T_{max})$. This continuity then further implies that there is a  constant $\delta>0$ such that $\|e(t)\|_\XX \leq R$ on $[0,T_0+\delta]$. But this contradicts Equation \ref{eq:sup_defn_T0}  that defines $T_0$ in terms of the supremum. 

Furthermore, it also must be the case that $T_{max}=\infty$. We have shown that $V(t)\leq V(0)$ for all $t\in [0,T_0]$. We have already proven that $\|e(t)\|_{\XX}<R$. Since $V(t)\leq V(0)$ for all $t\in [0,T_{max})$, the state $Z(t)$ is also bounded on this interval. Since $\|Z(t)\|\not \to  \infty$ as $t\to T_{max}$, it follows that $T_{max}$ cannot be finite. 

{Now the application of Barbalat's Lemma completes the proof.}
We can integrate $\dot{V}$ to find that 
\[
\frac{V(0)-V(t)}{\alpha} \geq  \int_0^t \sigma'(e^T(\tau)Pe(\tau))d\tau]\triangleq g(t).
\] 
 Since $V(t)$ is a nonincreasing function  that is bounded below, its limit $V_\infty$ as $t\to \infty$ exists, and the integral $g(t)$ on the right is consequently uniformly  bounded in time $t$. The integral $g(t)$ on the right hand side is consequently   nondecreasing and bounded above, so the limit of the integral on the right  exists as $t\to \infty$. This means that 
 \[
\frac{V(0)-V_\infty}{\alpha} \geq  \int_0^\infty \sigma'(e^T(\tau)Pe(\tau))d\tau.
\] 
 In addition, the integrand above is Lipschitz continuous. This follows since $\dot{e}(t)$ is continuous and uniformly bounded. The boundedness of $V$ ensures the boundedness of  $\tilde{K}_x$, $\tilde{K}_r$, $\tilde{\Theta}$ and $\tilde{f}$, and we have already established the boundedness of $e(t)$ for all time.  Or in other words, $e\in L^\infty(\RR^+,\XX)$, $\tilde{K}_x\in L^\infty(\RR^+,\RR^{n\times m})$, $\tilde{K}_r\in L^\infty(\RR^+,\RR^{m\times m})$, $\tilde{\Theta}\in L^\infty(\RR^+,\RR^{p\times m})$, and $\tilde{f}\in L^\infty(\RR^+,\vK)$. Thus, we have
\begin{align*}
    &\normRd{\dot{e}(t)}=\normRd{A_{ref} e + B\Lambda(-\tilde{K}_x^T x - \tilde{K}_r^T r +\tilde{\Theta}^T(t)\Phi( x(t))+ E_{x}\tilde{f}(t,\cdot))}, \\
    &\leq \|A_{ref}\| \|e(t)\|_\XX +  \|B\Lambda\|\biggl (\|\tilde{K}_x(t)\|_{\RR^{n\times m}} \|x(t)\|_\XX +\|\tilde{K}_r(t)\|_{\RR^{m\times m}} \|r(t)\|_\UU\\
    &+\|\tilde{\Theta}(t)\|_{\RR^{p\times m}}\|\Phi(x(t))\|_{\RR^{p}}  + \bar{k}\|\tilde{f}(t,\cdot)\|_\vK\biggr ) \leq \text{constant},
\end{align*}  
where we have used the fact that the operator-valued kernel is bounded on the diagonal and  $\|E_{x(t)}\|\leq \bar{\knl}$ for all $t\geq 0$.  Consequently, $e(t)$ is globally Lipschitz continuous. The composition of the locally Lipschitz continuous deadzone derivative $\sigma'$, which is globally Lipschitz continuous on $[0,\Delta]$,  and the globally Lipschitz continuous tracking error $e(t)$ yields an  integrand that is globally Lipschitz continuous. Since the integrand is therefore uniformly continuous, Barbalat's Lemma implies that the integrand converges to zero, and theorem is proved.

\end{proof}

\bigskip 

\begin{remark}
    The ultimate controller performance bound above takes the form 
    \[
    \limsup_{t\to \infty} \|x(t)-x_r(t)\|_\XX \leq C \sup_{\xi\in \overline{B_{\bar{R}}(0)}}\|E_\xi(I-\bm{\Pi}_N)f\|_\UU, 
    \]
    where $\overline{B_{\bar{R}}(0)}$ is the closed ball of radius $R$ in $\XX\triangleq \RR^n$. This form of the performance bound holds for any $f\in \vK$ and does not as of yet use any particular approximation properties of $\vK$. 
\end{remark}

The characterization of the approximation error for  uncertainty contained in the maneuver vRKHS $\vK_\calM$ can be now used directly in conjunction with the deadzone controllers in Theorem  \ref{th:smoothed_deadzone_AAO}.  To begin, however,  we start with a negative result of the type described in the introduction, one that illustrates how the curse of dimensionality can manifest in the nonparametric adaptive control setting.  

Suppose that the scalar-valued kernel $\knl$ defines RKHS $\sK(\XX,\RR)$, and to keep the notation simple we use the same symbol for the restriction  $\sK(\Omega,\RR)$ to the subset $\Omega\triangleq \overline{B_{\bar{R}}(0)}\subset \XX$.  We choose quasiuniform centers $\Xi_N\subset \Omega$ and define the associated finite dimensional subspaces
\begin{align*}
\sK_N\triangleq \text{span}\{\knl_{\xi_i}\ | \ \xi_i\in \Xi_N,1\leq i\leq N\}.
\end{align*}
It is well-known that  that the power function $\Pwr_{N}(x)$ of the subspace $\sK_N$ in $\sK(\Omega,\RR)$ can often be bounded above in the form 
\begin{align}
\Pwr_N(x)\leq C h^s_{\Xi_N,\Omega} \quad \quad  \text{for all } x\in \Omega, \label{eq:bound_power}
\end{align}
for a smoothness parameter $s>0$ that depends on the type of kernel. For example, polynomial powers, thin plate splines, Wendland functions, and Sobolev-Matern kernels all have this property for various choices of the smoothness parameter $s>0$. See Table of \cite{wendland}, Table 1 of \cite{schaback94}, or the discussion in \cite{kurdilaBook}. 
\begin{corollary}
\label{cor:curse}
    Let the scalar-valued kernel $\knl:\XX \times \XX$ be defined as above, so that the upper bound on the power function in Equation \ref{eq:bestbounds} holds for a smoothness integer $s\geq 1$. Define the diagonal operator-valued power function $\Knl\triangleq \knl I_m$ where $I_m$ is the identity operator on $\RR^m$, so that $\vK\triangleq \sK^m$, and let $\Xi_N\subset \Omega \triangleq \overline{B_{\bar{R}}(0)}$ be a quasiuniform set of centers in $\Omega$.  Then to achieve a target ultimate tracking error 
    \[
    \limsup_{t\to \infty}\|x(t)-x_r(t)\|_\XX \leq \epsilon, 
    \]
    the performance bound in Theorem \ref{th:smoothed_deadzone_AAO} implies that we must choose the number of centers 
    \[
    N=N(\Xi_N,\Omega) \geq \frac{1}{\epsilon^{n/s}}. 
    \]
\end{corollary}
\begin{remark}
    As noted in the introduction, this bound exhibits the classical curse of dimensionality as the dimension $n$ of the state space increases. While we have established this result using approximations from RKHS, it should also be noted that similar results follow for popular choices where the space of approximants are defined in terms of finite elements or splines. 
\end{remark}

Now we begin our analysis of how this basic performance estimate can be improved through the use of the maneuver vRKHS defined in this paper. 
We start with the following, which is the most general  and treats the case of a general, possibly nondiagonal,  operator-valued kernel. 
\begin{corollary}
\label{cor:general_R_M}
    Suppose that the general operator-valued kernel $\Knl$ is used to define the maneuver vRKHS $\Knl_\calM$ as described in Theorem \ref{th:doublingtrick}. Then the smoothed deadzone method described in Theorem \ref{th:smoothed_deadzone_AAO} has an ultimate bound on the tracking error that is given by 
    \begin{align*}
        \limsup_{t\to\infty}\|x(t)-x_r(t)\|_\XX  \leq C \sup_{\xi\in \calM}\bar{\Pwr}_{\vR_N}(\xi)\|r\|_{\vR_\calM}\|v\|_{L^2(\calM,\YY)} 
    \end{align*}
    for a constant $C>0$, where the uncertainty $f=\Ext_\calM r$ for the unique $r\in \vR_\calM$ that satisfies the regularity condition  $r=Lv$ for some $v\in L^2_\mu(\calM,\YY)$. 
\end{corollary}

The following corollary establishes that the maneuver vRKHSs enable computational complexity estimates that scale like those described in the introduction.
\begin{corollary}
\label{cor:diagonal_R_M}
 Suppose that the diagonal operator kernel $\Knl\triangleq \knl I_m$ is  defined in terms of   the scalar-valued kernel $\knl$ that either satisfies the regularity condition in Equation \ref{th:diagonal_C2s}  or the algebraic decay conditions in Theorem \ref{th:many_zeros}. Then the nonparametric adaptive control method defined by the family $\{\Delta_N\}_{N\geq N_0}$  in Theorem \ref{th:smoothed_deadzone_AAO} satisfies the ultimate performance bound 
    \[
    \limsup_{t\to \infty} \|x(t)-x_r(t)\|_\XX \sim h^{\bar{s}}_{\Xi_N,\calM}
    \]
    for all functional uncertainty $f$ in the maneuver vRKHS $\vK_\calM\subseteq \vK$ of globally defined functions on the state space $\XX$. 
    If the family of centers $\{\Xi_N\}_{N\geq N_0}$ is quasiuniformly distributed in the manifold $\calM$, the bound above implies that  the number of centers $N$ required to achieve the ultimate target tracking error $\epsilon>0$ scales like 
    \[
    N\triangleq N(\calM)\sim \frac{1}{\epsilon^{\ell/\bar{s}}}.
    \]
\end{corollary}

\section{Conclusions and Future Work}
\label{sec:conclusions}
This paper has introduced a general approach for the construction of infinite dimensional vRKHSs for the representation of functional uncertainties in nonparametric adaptive control problems.  We   refer to them as maneuver vRKHS spaces. The design of the maneuver vRKHSs is carried out by exploiting information that is ordinarily readily available in many model reference adaptive control problems that seek to track a reference system. We suppose that the reference system to track ultimately approaches  a compact, connected, smooth, $\ell$-dimensional Riemannian submanifold $\calM$ that is regularly embedded in the full state space $\XX\triangleq \RR^n$.  The approach is motivated by the intuition that ultimately driving the tracking error to be small is made possible when the approximation error  for the uncertainty is small over the set that ultimately  supports the dynamics. 

To study the computational complexity of an adaptive feedback controller for a system with functional uncertainty in a maneuver space, we first show that, if we are not careful in the selection of the vRKHS for representations of uncertainty,  even some approximation theory asymptotically optimal (AA0) nonparametric adaptive control schemes can exhibit poor scaling as the state space dimension grows. However, by designing  a maneuver space $\vK_{\calM}$, we establish that the computational complexity of the adaptive control satisfies a better scaling bound that depends on the dimension of the embedded submanifold $\ell$ and reduced smoothness $\bar{s}$ of functions in the maneuver space $\vK_\calM$.

Despite these promising features, the research of this paper suggests a number of important open problems that can be addressed in future work. The current design process selects the dimension $\ell$ of the embedded limiting submanifold and the reduced smoothness $\bar{s}>0$. But finite dimensional approximations of used in the synthesis of practical controllers is based on selection of centers that live on the manifold. While we expect that this is appropriate for good approximations of the uncertainty for larger values of $t$, and good ultimate performance, it is not clear that approximation errors may be larger during the transient regime when the trajectory is relatively far from the embedded manifold. One important future research topic would be the development  of data-driven addition and/or deletion  of centers along the trajectory: ultimately these would accumulate at or near the limiting manifold. This would entail the design of maneuver spaces that are designed over subsets of the state space that contain the limiting manifold. 

\clearpage 
\section{Appendices}

\subsection{Feature Mappings, Operators,  and Spaces for vRKHS} 
\label{sec:vector_feature}
Various theorems can be found in the literature that study and characterize feature spaces for scalar or vector-valued RKHS spaces. See \cite{scholkopf2002learning} for an in-depth discussion of feature operators in a scalar-valued  RKHS setting or \cite{carmeli10} for their study in  vRKHSs.   The theorems in this latter reference provide  one popular way to construct very general vRKHS that contain functions $f:X\to Y$ defined over a subset $X$ that take values in a Hilbert space $Y$.  The discussion that follows  uses  many of the properties  feature operators, feature mappings, and  feature spaces as described  in Proposition 1 from \cite{carmeli10}. This proposition  requires  that two fundamental assumptions  hold on the set $X$ that determines the domain over which functions are defined and on the Hilbert space $Y$. 
\begin{enumerate}
    \item[A1)] The set  $X$ is a locally compact, second countable topological space.
    \item[A2)] The Hilbert space $Y$ is separable.
\end{enumerate}
  While the topological condition on $X$ is rather abstract, it poses no real restriction on the problems of interest studied in this paper. In our applications $X$ is either  a Euclidean space $\RR^d$ for some $d\geq 1$, a compact Riemannian manifold, or a subset of these  sets. Likewise, in the only cases considered here, the finite dimensional  Hilbert space $Y\triangleq \YY\triangleq \RR^m$  satisfies assumption $(A2)$.

\begin{theorem}[\cite{devito06}, Proposition 1]
\label{th:vector_feature}
Let (A1) and (A2) hold, let $U$ be a Hilbert space, and  $\Psi:X \rightarrow \calL(Y,U)$. The feature  operator 
$
F:U\rightarrow \cF(X,Y)$ defined by 
$$
(Fu)(x)=\left (\Psi(x) \right)^*u \quad \quad \text{ for all } x\in X, u\in U, 
$$
is a partial isometry $F:U\rightarrow \vK_\Psi$  onto the $Y$-valued RKHS space $\vK_\Psi\triangleq \Range{F}$  that is generated by the admissible operator-valued kernel
\begin{align}
\Knl_\Psi(x_1,x_2):=\left (\Psi(x_1) \right )^*\Psi(x_2) \quad \quad \text{ for all } x_1,x_2\in X. \label{eq:kernel_Kpsi}
\end{align}
The  operator $\Pi_{U_I}\triangleq F^*F:U \rightarrow U$ is the $U-$orthogonal projection onto the initial space 
$$
U_I\triangleq  xll{F}^\perp=\overline{
\text{span}\left \{
\Psi(x)y\ | \ x\in X, y\in Y
\right \}
}
$$
and 
\begin{equation}
\|f\|_{\vK_\Psi}=\text{inf} \left \{
\|u\|_U\ | \ f=Fu, u\in U
\right \}. \label{eq:infdefnHPsi}
\end{equation}
\end{theorem}

The following  observations are also useful when we apply the above theorem in this paper.
\begin{enumerate}
\item It is also important to note that Lemma 2 of \cite{zhang2012refinement} establishes that the inner product on $\vK_\Psi$ is given by
\begin{align*}
   \left ( Fu,Fv\right)_{\vK_\Psi}&=\left  (\Psi(\cdot)^*u,\Psi(\cdot)^*v\right)_{\vK_\Psi} \\
   &\triangleq \left ( \Pi_{U_I} u, \Pi_{U_I} v\right)_U \quad \quad \text{ for all } u,v\in U,
\end{align*}
where $\Pi_{U_I}$ is the $U$-orthogonal projection of $U$ onto 
\[
U_I\triangleq \overline{\text{span}\left \{ \Psi(x)y \ | \ x\in X, y\in Y\right \} }\subseteq U.
\]
\item The infimum in Equation \ref{eq:infdefnHPsi} is actually achieved since the feature map $F$ is a partial isometry.  In fact for any $f\in \vK_\Psi$ we have  
\[
\|f\|_{\vK_\Psi}=\|F^*f\|_{U}. 
\]
This follows since whenever we have $f=Fu$, we can write
\begin{align*}
\|u\|^2_U&=\|\Pi_{U_I}u\|^2_U + \|(I-\Pi_{U_I})u\|_U^2 \\
&=\|F^*Fu\|_{U}^2 + \|(I-\Pi_{U_I})u\|_U^2 \\
&=\|F^*f\|_{U}^2 + \|(I-\Pi_{U_I})u\|_U^2 \geq \|F^*f\|_U^2.
\end{align*}
Also, since $FF^*$ is the identity on $\vK_\Psi$  by Theorem \ref{th:partialisometry},   the definition of $\|f\|_{\vK_\Psi}$ enables the upper bound  
\[
\|f\|_{\vK_\Psi}=\text{inf} \left \{
\|u\|_U\ | \ f=Fu, u\in U
\right \}\leq \|F^*f\|, 
\]
because $f=F(F^*f)$ and the choice $u=F^*f$ is in the set on the right over which the infimum is computed. See also Theorems 2.2.3 and 2.3.4 of \cite{wittwar2022approximation} which can be construed as  some specific cases where the minimum in the infimum above is realized in the vRKHS setting. In the scalar-valued RKHS setting, see also Theorem 6 of \cite{berlinet}, or Equation 2.235 of Theorem 2.36 and Equation 2.240 of Theorem 2.37 in \cite{saitoh}. Equation 2.240 is precisely the statement that the corresponding  infimum in Equation 2.230 of \cite{saitoh} in the scalar-valued setting is achieved. 
\item 
It can be verified directly that $\Knl$ in Equation \ref{eq:kernel_Kpsi} is the reproducing kernel of $\Knl_\Psi$, and this calculation can be useful in applications of the theorem. We have 
\begin{align*}
    \Knl_{\Psi,x}(\cdot)y&\triangleq \Knl_\Psi(\cdot,x)y=F\psi(x)y \quad \quad \text{ for all } x\in X, y\in Y.
\end{align*}
We know that any  $h\in \Knl_\Psi$ has a  representation $h=Fu$ for some $u\in U_I$, so that 
\begin{align*}
    \left (h,\Knl_{\Psi,x}y\right)_{\Knl_\Psi} &=
    \left ( (F|_{U_I^\perp})^{-1}Fu,(F|_{U_I^\perp})^{-1} F\psi(x)y\right)_U,\\
    &=\left (\Pi_{U_I}u,\Pi_{U_I}\Psi(x)y) \right)_{U}=\left ( u, \Psi(x)y\right)_U,\\
    &=\left ( \Psi^*(x)u,y \right)_\YY,\\
    &=\left (h(x),y\right)_\YY=\left ( E_x h,y\right)_\YY,
\end{align*}
for all $h\in \Knl_\Psi,y\in \YY$. 
\end{enumerate}

While this theorem is abstract, some intuition regarding its structure can be gleaned by considering Figure \ref{fig:vector_feature}.
%\begin{figure}
%    \centering
%    \includegraphics[width=4in]{featuremapping.pdf}
%    \caption{Embedding by the Feature Operator $F$}
%    \label{fig:vector_feature}
%\end{figure}
The feature map $F:U\rightarrow \vK_\Psi$ is used to split the Hilbert space $U$ into two closed subspaces. The large space $U$ is decomposed into the $U-$orthogonal sum of $\Null{F}$ and its orthogonal complement $U_I\triangleq \Null{F}^\perp$.  The feature map $F$ is onto the native space $\vK_\Psi$ guaranteed by the theorem, and it is a partial isometry. The restriction of the feature map to $\Null{F}^\perp$ is an isometry onto $\vK_\Psi$. In this sense, the native space $\vK_\Psi$ can be identified with a closed subspace of $U$.

The above theorem is used in a host of situations, and the interested reader is referred to \cite{carmeli10,devito06,zhang2012refinement} for some concrete examples. A canonical example of the use of the above theorem is given in the next section, where we choose $X=\XX=\RR^n$. This choice  is used to relate subspaces of vector-valued functions over $\XX$  and vector-valued functions that are restricted to a subset $\Omega\subseteq \XX$. 
%
%%  subspaces generated by subsets and restrictions
%
\subsection{Subspaces Generated by a Subset $\Omega\subset \XX$ or  Restrictions to $\Omega\subseteq \XX$}
In our studies of spaces generated by operator-valued kernels, we often follow a standard practice. We choose a convenient or well-known vRKHS that contains functions defined on all of $\XX$, and then we construct associated vRKHS that are defined in terms of some subset $ \Omega \subseteq \XX$. Since we choose $X=\XX\triangleq \RR^n$ and $y=\YY\triangleq \RR^m$, the assumptions (A1) and (A2) hold.

Suppose now we are given a $\YY-$valued native space $\vK$ of functions defined over $\XX$ in terms of the admissible operator-valued kernel $\Knl:\XX\times \XX\rightarrow \calL(\YY)$. We fix a subset $\Omega\subseteq \XX$ and  define the sets of functions
\begin{align}
    \vK_\Omega:&=\overline{\text{span}
    \left \{\Knl_\omega y\ | \ \omega\in \Omega, y\in \YY \right \}}\subseteq \vK, \label{eq:vHS} \\
    \vR_\Omega:&=\left \{ g:\Omega \rightarrow \YY\ | \ g=T_\Omega f= f|_{\Omega}, f\in \vK \right \}=\Range{T_\Omega}, \label{eq:vRS} \\
    \vZ_\Omega:&=\left \{f\in \vK \ | \
    T_\Omega f=f|_\Omega =0\right \} =\Null{T_\Omega}. \label{eq:vZS}
\end{align}
In the definition above of $\vK_\Omega$, the closure on the  linear span is taken with respect  the norm on the space $\vK$, so that $\vK_\Omega$ is a closed subspace of $\vK$. The operator $T_\Omega$ in the definition of $\vR_\Omega$ is the  trace or restriction operator given by 
\[
(T_\Omega f)(\omega)=f(\omega)\in \YY \quad \quad \text{ for all } \omega\in \Omega. 
\]
It  is emphasized that $\vK_\Omega$ contains functions over $\XX$, while $\vR_\Omega$ contains functions only defined on $\Omega\subseteq \XX$. 
\begin{theorem}
\label{th:vectorZSHS}
Suppose that  the subset $\Omega\subseteq \XX$  and  $\Knl:\XX\times \XX\rightarrow \calL(\YY)$ is an admissible operator-valued kernel that defines the $\YY-$valued RKHS space $\vK$ of functions defined over $\XX$. The function spaces $\vK_\Omega$ and $\vZ_\Omega$ defined in Equations \ref{eq:vHS} and \ref{eq:vZS} comprise the  $\vK$-orthogonal decomposition
\[
\vK=  \vK_\Omega \oplus \vZ_\Omega,
\]
and the  reproducing kernel of $\vK_\Omega$ is given by 
\begin{align}
    \Knl_\Omega(x_1,x_2)&=E_{x_1}\bm{\Pi}_\Omega E_{x_2}^* = \Knl_{x_1}^* \Pi_\Omega \Knl_{x_2} \quad \text{ for all } x_1,x_2 \in \XX,  \label{eq:K_Omega_kernel}
\end{align}
where $E_x=(\Knl_x)^*$  is the evaluation operator on $\vK$ at $x\in \XX$ and $\bm{\Pi}_\Omega$ is the $\vK$-orthogonal projection of $\vK$ onto $\vK_\Omega$. 
The space of restrictions $\vR_\Omega$ is a $\YY-$valued native space that is induced by the restricted operator kernel $\Rnl_\Omega: \Omega \times \Omega \rightarrow \calL(\YY)$ given by $\Rnl_\Omega\triangleq \Knl|_{\Omega\times \Omega}$,  so that 
\[
\Rnl_\Omega(\omega_1,\omega_2):=\Knl(\omega_1,\omega_2) \quad \text{ for all } \omega_1,\omega_2\in \Omega.
\]
The norm induced by this kernel on $\vR_\Omega$ is equivalent to the expression
\begin{align}
\|f\|_{\vR_\Omega}&\triangleq \inf \left \{
\|g\|_{\vH} \ | \ f=T_\Omega g=g|_\Omega, g\in \vK
\right \}, \label{eq:equiv_RS} \\
&
=\min \left \{
\|g\|_{\vH} \ | \ f=T_\Omega g=g|_\Omega, g\in \vK
\right \}.
\end{align}
The adjoint operator $T^*_\Omega$ defines a canonical extension operator via 
\[
\Ext_\Omega\triangleq T^*_\Omega:\vR_\Omega \to \Range{\Ext_\Omega}=\vK_\Omega \subseteq \vK. 
\]
The restriction and extension operators 
\begin{align*}
    T_\Omega|_{\vK_\Omega}&:\vK_\Omega \to \vR_\Omega, \\
    \Ext_\Omega&:\vR_\Omega \to \Range{\Ext_\Omega}=\vK_\Omega,
\end{align*}
are onto isometries. 
\end{theorem}
\begin{proof}
This theorem is proven by applying Theorem \ref{th:vector_feature} a couple of times,  once for the space $\vR_\Omega$ and once for the space $\vK_\Omega$.  

\noindent \underline{The Feature Operator  $F:\vK\to \vR_\Omega$:}
We choose  $U$  to be $\vK$, the subset $\Omega\subseteq \XX$, and  
\begin{align*}
    \Psi(\omega)&:=\Knl_\omega\in \calL(\YY,\vK) \quad &\text{ for all }  &\omega\in \Omega,& \\
    (Fh)(\omega)&:=\underbrace{(\Knl_\omega)^*}_{\Psi(\omega)^*\in \calL(\vK,\YY)}h \quad &\text{ for all } &h\in \vK.&
\end{align*}
With these choices, the feature map $F$ is precisely the trace operator $T_\Omega$  since 
\[
(Fh)(\omega)=(\Knl_\omega)^*h=E_\omega h=h(\omega) \quad \quad \text{ for all } \omega\in \Omega, 
\]
that is, $Fh=T_\Omega h$.
In this case the operator-valued kernel $\Rnl_\Omega:\Omega \times \Omega \to \calL(\YY)$ is just the restriction of the operator $\Knl:\XX \times \XX \to \calL(\YY)$ since 
\[
\Rnl_\Omega(\omega_1,\omega_2)\triangleq \Psi(\omega_1)^*\Psi(\omega_2)=\Knl_{\omega_1}^*\Knl_{\omega_2}=\Knl(\omega_1,\omega_2)\quad \quad \text{ for all } \omega_1,\omega_2\in \Omega. 
\]
The feature mapping theorem then implies that 
\begin{align}
    \vK=\vK_\Omega \oplus \vZ_\Omega \label{eq:K_orthog_decomp_1}
\end{align}
since $\vZ_\Omega=\Null{T_\Omega}$ and 
the initial space $U_I\subseteq \vK$ is given by 
\begin{align*}
U_I&=\overline{\text{span}\{\Psi(\omega)y\ | \ \omega\in \Omega,y\in \YY\}}\\
&=\overline{\text{span}\{\Knl_\omega y\ | \ \omega\in \Omega,y\in \YY\}} =\vK_\Omega,
\end{align*}
where the closure above is taken in the norm of $\vK$.

\noindent \underline{The Feature Operator  $F:\vK\to \vK_\Omega$:} Alternatively, we choose $U=\vK$, the subset $\Omega\subseteq \XX$, and  define the feature operator to be the $\vK$-orthogonal projection $F=\bm{\Pi}_\Omega:\vK\to \vK_\Omega$. In this case we have 
\begin{align*}
    (Fh)(x)&:=\underbrace{E_x \Pi_\Omega}_{\Psi(x)^*\in \calL(\vK,\YY)} h \quad &\text{ for all } &h\in \vK,x\in \XX.&
\end{align*}
The theorem then guarantees that $\vK_\Omega$ is a vRKHS with operator kernel 
\[
\Knl_\Omega(x_1,x_2)\triangleq \Psi(x_1)^* \Psi(x_2)=E_{x_1}\Pi_\Omega \Pi_{\Omega}^* E_{x_2}^* \in \calL(\YY) \quad \quad \text{ for all } x_1,x_2 \in \XX. 
\]
It is straightforward to verify that the definition of the operator-valued kernel  $\Knl_\Omega$ above does indeed satisfy the reproducing property over $\vK_\Omega$ since 
\begin{align*}
    \left (\Knl_{\Omega,x}y,h \right)_{\vK}&=  \left (\bm{\Pi}_\Omega\Knl_{x}y,h \right)_{\vK} 
    =  \left (\Knl_{x}y,\bm{\Pi}_\Omega h \right)_{\vK},\\
    &=   \left (\Knl_{x}y,h \right)_{\vK}=\left (y,E_xh \right)_\YY,\\
    &=(y,h(x))_\YY, \quad \quad \text{ for all } x\in X, y\in Y, h\in \vK_\Omega.
\end{align*}
Since $F=\bm{\Pi}_\Omega$ and $\Null{F}=(I-\bm{\Pi}_\Omega)\vK$, in this case, the  feature mapping theorem also implies  that 
\begin{align}
    \vK&=\bm{\Pi}_\Omega(\vK)\oplus (I-\bm{\Pi}_\Omega)(\vK) \notag \\
    &=\vK_\Omega \oplus \vZ_\Omega. \label{eq:K_orthog_decomp_2}
\end{align}
It is possible to interpret the two applications  of the feature mapping theorem above in Equations \ref{eq:K_orthog_decomp_1} and \ref{eq:K_orthog_decomp_2} as a proof that 
\[
\vZ_\Omega = (I-\bm{\Pi}_\Omega)(\vK).
\]
\end{proof}

\begin{remark}
    In the above constructions, note that $\vR_\Omega =T_\Omega(\vK_\Omega)=T_\Omega(\vK)$. We also can relate the operator-valued kernels $\Rnl_\Omega$ and $\Knl_\Omega$. We have 
    \[
    \Rnl_\Omega = \Knl_\Omega|_{\Omega \times \Omega}=\Knl|_{\Omega \times \Omega}. 
    \]
    This holds since we can compute directly 
    \begin{align*}
\Knl_\Omega(\omega_1,\omega_2)
&=E_{\omega_1}\bm{\Pi}_\Omega E^*_{\omega_2}=E_{\omega_1}
E^*_{\omega_2}\\
&=\Knl(\omega_1,\omega_2)
=\Rnl_\Omega(\omega_1,\omega_2) \quad \quad \text{ for all } \omega_1,\omega_2 \in \Omega \subset \XX. 
    \end{align*}
    While the above identity implies that the restricted kernels $\Rnl_\Omega=\Knl|_{\Omega\times \Omega}=\Knl_\Omega|_{\Omega\times \Omega}$ are identical, it is not true that $\Knl=\Knl_\Omega$ over all of $\XX\times \XX$. 
\end{remark}

\begin{remark}
    The analysis in this theorem derived the form of the operator kernel $\Knl_\Omega$ of the vRKHS $\vK_\Omega$ that is generated by the set $\Omega\subset \XX$. However, an entirely analogous application of the argument in the second half of the proof gives a more general result. If $\bm{\Pi}_{\vU}$ is the $\vK$-orthogonal projection onto the closed subspace $\vU\subseteq \vK$, then the subspace $\vU\triangleq \vU(\XX,\YY)$ is a vRKHS for the operator-valued kernel 
    \begin{align}
        \Unl(x_1,x_2)\triangleq E_{x_1} \bm{\Pi}_{\vU} E_{x_2}^* \in \calL(\YY)\quad \quad \text{ for all } x_1,x_2\in \XX. \label{eq:operator_kernel_U}
    \end{align}
\end{remark}

The operators $\Ext_\Omega\triangleq T_\Omega^*$ and the $T_\Omega$ are extremely useful. 
They enable passing  between  the spaces $\vK=\vK(\XX,\YY)$ or $\vK_\Omega=\vK_\Omega(\XX,\YY)$ that contain ``global'' functions supported on all of $\XX$ and the space $\vR_\Omega=\vR_\Omega(\Omega,\YY)$ that contains ``local'' functions supported on just $\Omega\subset \XX$. It should be emphasized that the  general study of extension and restriction operators in various function spaces can be a  delicate undertaking, see \cite{brudnyi2011methods}.  It can consequently be somewhat surprising that  the situation in a vRKHS is relatively simple. There always exist canonical bounded linear  extension and restriction operators that relate the spaces $\vK$, $\vK_\Omega$ of globally defined functions  to the space of restrictions   $\vR_\Omega$ \textit{for any subset $\Omega\subset \XX$ without consideration of any regularity properties  $\Omega$ may or may not have}. 

This fact has been pointed out by other authors. For comparison, see Lemma 4 in \cite{fuselier2012} or Section 10.8 of \cite{wendland}, both of which study the setting of scalar-valued RKHS.   For example, \cite{fuselier2012} provides  a different proof of the existence of a canonical linear extension operator for spaces of restrictions $\mathcal{R}_\Omega$ of  scalar-valued functions in RKHS $\sK=\sK(\XX,\RR)$. In  reference \cite{fuselier2012}  the extension operator is defined as the an extension by continuity of the  mapping 
\begin{align*}
    \Ext_\Omega: \sum_{i=1}^N  \rnl_{\Omega,\xi_i}\theta_i \mapsto \sum_{i=1}^N \knl_{\xi_i}\theta_i, 
\end{align*}
where $\rnl_\Omega=\knl|_{\Omega\times \Omega}$ is the restricted scalar-valued kernel that defines $\sR_\Omega=T_\Omega(\sH(\RR^n,\RR))$, the centers $\xi_i\in \Xi_N\subset \Omega$, and $\theta_i\in \RR$ are real coefficients. 
While the proof in \cite{fuselier2012} does not use feature mappings, the final linear extension operator for any subset $\Omega$ is the same as that derived in this paper.  For the vector-valued setting we could alternatively define the  operator as the extension by continuity of the mapping 
\begin{align*}
    \Ext_\Omega: \sum_{i=1}^N  \Rnl_{\Omega,\xi_i}\Theta_i \mapsto \sum_{i=1}^N \Knl_{\xi_i}\Theta_i, 
\end{align*}
where $\Theta_i\in \YY$. 

Before concluding this short section, we illustrate one more way to interpret the extension operator. 
Since the trace operator $T_\Omega:\vK \to \vR_\Omega$ can be interpreted as a feature mapping onto $\vR_\Omega$, we know that the operator $T_\Omega|_{\vH_\Omega}:\vK_\Omega \to \vR_\Omega$ is an isometry, so that its inverse $\left ( T_\Omega |_{\vK_\Omega}\right)^{-1}:\vR_\Omega \to \vK_\Omega$ exists and 
\[
\left \|(T_\Omega|_{\vH_\Omega})^{-1} \right\| = 1. 
\]
Thus, we can always define a bounded linear  extension operator 
$$
\overline{\Ext}_{\Omega}\triangleq (T_\Omega|_{\vK_\Omega})^{-1}:\vR_\Omega \to \vK_\Omega\triangleq \Range{T_\Omega} \subset \vK. 
$$
As we show below, again using the feature mapping Theorem \ref{th:vector_feature}, the operator $\bar{\Ext}_\Omega$ is in fact given by $\Ext_\Omega\triangleq T_\Omega ^*$. 

Recall that in the comments following Theorem \ref{th:vector_feature}, 
for a feature mapping with $\Range{F}= \vK_\Psi$, we can write the inner product as   
\[
 (h_1,h_2)_{\vK_\Psi} \triangleq \left ((F|_{U_I})^{-1}h_1,(F|_{U_I})^{-1}_2 h_2\right )_U \quad \quad \text{ for all } h_1,h_2 \in \vK_\Psi.
\]
For the specific case at hand, when we choose $U=\vK$, $\vK_\Psi=\vR_\Omega$ and the feature operator $F=T_\Omega$, this defines the inner product 
$$
\left ( r_1,r_2\right )_{\vR_\Omega} \triangleq \left ( (T_\Omega |_{\vK_\Omega})^{-1} r_1, (T_\Omega |_{\vK_\Omega})^{-1} r_2\right )_\vK \quad \quad \text{ for all } r_1,r_2\in \vR_\Omega. 
$$
Finally, we directly compute  
\begin{align*}
    \left ( T_\Omega h,r\right )_{\vR_\Omega}&=
\left ( (T_\Omega |_{\vH_\Omega})^{-1} T_\Omega h, (T_\Omega |_{\vK_\Omega})^{-1} r\right )_\vK\\
&=\left ( \Pi_\Omega h, (T_\Omega |_{\vH_\Omega})^{-1} r\right )_\vH = \left (  h, (T_\Omega |_{\vH_\Omega})^{-1} r\right )_\vH\\
&= \left (  h, T_\Omega ^* r\right )_\vK, \quad \quad \text{ for all } h\in \vK,r\in \vR_\Omega.
\end{align*}
Therefore, 
$$
\overline{\Ext}_\Omega = (T_\Omega|_{\vK_\Omega})^{-1}= T^*_\Omega=\Ext_\Omega.
$$

\bibliographystyle{abbrv}
\bibliography{all_references1,all_references2,References1}
\end{document}